\documentclass{article}

\usepackage{graphicx}
\usepackage{dcolumn}
\usepackage{bm}

\usepackage[utf8]{inputenc}
\usepackage[T1]{fontenc}
\usepackage{mathptmx}
\usepackage{etoolbox}
\usepackage{fdsymbol}

\usepackage{doi}
\usepackage{graphicx}
\usepackage[giveninits=true, backend=biber]{biblatex}
\usepackage{hyperref}
\usepackage{url}

\newcommand{\pun}{\mathbin{\largesquare}}
\newcommand{\pdeux}{\mathbin{\largesquare\hspace{-7.5pt}\times}}
\newcommand{\pcinqa}{\mathbin{\bigcirc}}
\newcommand{\pcinqb}{{\largetriangleup}}
\newcommand{\pcinqc}{{\Diamond}}
\newcommand{\psix}{\largetriangledown}

\title{Structure analysis of the {\it Lorenz--84} chaotic attractor}
\author{
    \begin{minipage}{\textwidth}
    \centering
    M. Rosalie\textsuperscript{1}, S. Mangiarotti\textsuperscript{2}\\
    \bigskip
    {\small
    1. Université de Perpignan Via Domitia, CNRS, Laboratoire Génome et Développement des Plantes UMR-5096, F-66860 Perpignan, France\\
    2. CESBIO - Centre d'études spatiales de la biosphère, 18, Av Edouard Belin, 31401 Toulouse, France
}
\end{minipage}
}%

\begin{document}

\maketitle

\begin{abstract}
The structure of the {\it Lorenz--84} attractor is investigated in this study. Its dynamics belonging to weakly dissipative chaos, classical approaches cannot be used to analyze its structure. The color tracer mapping is introduced for this purpose and used to extract the three-dimensional structure of the attractor. The analysis shows that the attractor is a non trivial case of toroidal chaos: it is organized around a period--2 cavity. 
Moreover, the structure reveals a new mechanism generating chaos in the attractor: a multidirectional stretching. 
The attractor structure is then artificially represented on a two-dimensional branched manifold and its validation performed using a set of periodic orbits previously extracted.
\end{abstract}

\maketitle

{\bf 
Weakly dissipative chaos was thought to be rare since the early 1960s. Indeed, relatively few cases were found until 2010. During the last fifteen years, many cases were directly obtained from observations of the real world, for crop cycles in semi-arid regions and in eco-hydrology. This type of chaos cannot be analyzed using the usual topological techniques because these were developed for strongly dissipative attractors which structure is locally flat almost everywhere, as it is the case for the most famous {\it Lorenz--63} or {\it Rössler--76} attractors. On the contrary, weakly dissipative chaotic attractors are thick and therefore difficult to reduce to flat structure – a template. In this study, the first weakly dissipative chaotic attractor discovered by Edouard N. Lorenz in 1984 is investigated. Its structure is analyzed revealing a non trivial type of toroidal chaos and a stretching mechanism enabling to develop weakly dissipative chaos.
}

\section{Introduction}

Chaotic attractors are important in science because they powerfully show the possibility to have together determinism and unpredictability. If we put the chaotic maps aside, most of the earliest chaotic attractors found were strongly dissipative, as it is the case for the paradigmatic Lorenz \cite{Lrz63} and Rössler \cite{Ross76} attractors, and for many other early chaotic attractors \cite{ Rossler-Ortoleva-1978, PikoRabi-1978, Shaw-1981, Sprott-1994}. For such types of three-dimensional chaos, the Lyapunov spectrum is such as the positive exponent is much smaller in amplitude that the negative one on average ($\lambda_1 << |\lambda_3|$, with $\lambda_2 = 0$ corresponding to the direction of the flow), so that their Kaplan-Yorke dimension is larger but close to two (typically such as $2 < D_{KY} < 2.1$). Being strongly dissipative, these attractors are all very flat locally, and then characterized by a thin Poincaré section and a thin first return map as well. Their structure can then be commonly represented by a two-dimensional branched manifold called a template.
Topology of chaos has proven to be a very powerful approach to characterize such dynamics \cite{Mindlin1990, Gilmore1998, GiLef2002} that provides a detailed and non ambiguous description of the dynamics. The techniques have been developed for these systems enabling to characterize most of the three-dimensional systems \cite{Let1995, LetGil2013} except two types of chaotic attractors found more challenging: (i) the toroidal and (ii) the weakly dissipative ones.
Few approaches have addressed this problem.
One method has been introduced to precisely identify the best partition between branches in the first return maps\cite{Plumecoq_2000}. Introduced in the late 1990s, a homology based technique \cite{Sciamarella1999, Sciamarella2001} has proven to be very promising for the analyses in dimension three and higher \cite{Charo2022} but has not been applied yet to such weakly dissipative cases.
Recently, a template has been proposed for the toroidal chaotic attractor \cite{Man2021} solution to the Deng system \cite{Deng1994}. However, the representation of weakly dissipative attractors remains challenging.
The first weakly dissipative attractor ($D_{KY} \sim 2.39$) was discovered by Edward N. Lorenz in 1984 \cite{lorenz1984irregularity}. This attractor is solution to a three-dimensional system introduced by him to approximate the dynamics of the global scale atmosphere with three variables, only. Several other weakly dissipative systems were published, one for the dynamics of lasers \cite{Wieczorek1999} ($D_{KY} \sim 2.76$) as well as a modified version of the Lorenz—84 system \cite{Chlouv2005} ($D_{KY} \sim 2.54$).
In the early 2010, weakly dissipative attractors ($2.68 < D_{KY} < 2.75$) were directly extracted from observational time series for the cycles of cereal crops in North Morocco \cite{Man2012,Man2014} suggesting that this type of chaos was not just theoretical, but may also take place in the real world, conferring them a practical interest. This result was confirmed when analyzing other provinces in North Morocco for which many other cases of weakly dissipative chaos ($2.28 < D_{KY} < 2.80$) were obtained again. \cite{Man2023} A case of weakly dissipative chaos was also obtained for the dynamics of earthworms in tropical climatic conditions \cite{ManFu2021}. All these results reinforce the interest for the development of new approaches dedicated to the structural analysis of weakly dissipative attractors.

Several works have been dedicated to the Lorenz--84 system. Its bifurcation diagram was analyzed in detail \cite{broer2002bifurcations}. It was used to study the synchronization of weakly dissipative systems \cite{Sendi2017}. However, its attractor remains poorly known. The present study aims to characterize its structure. Due to its thickness, its analysis remains challenging. The methodology used to study its structure is presented in the next section 2. 
The third section details the transformations used to start performing topological characterization with numerical computation of liking number between periodic orbits.
The fourth section is dedicated to the extraction and analysis of its skeleton.
The next section synthesizes the results by providing a template for the Lorenz--84 attractor that permits to reproduce the linking numbers obtained numerically and to describe its topological structure as a reduced template.
The main conclusions are then presented in the last section.

\section{Methodology}
\label{sec:methodology}

\subsection{Unstable Periodic Orbit Extraction and characterization}

\begin{figure}[htb]
    \centering
    \includegraphics[width=.45\textwidth]{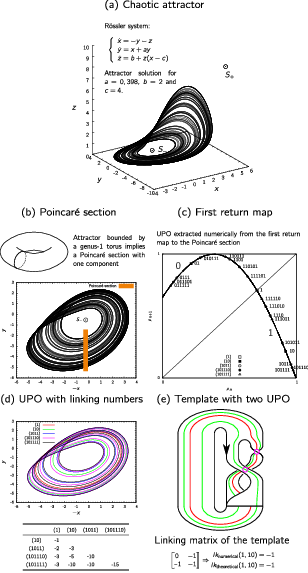} 
    \caption{
        Topological characterization methodology applied to the Rössler attractor \cite{Ross76} using reproducible methodology.
        $lk()$ refers to linking number between unstable periodic orbits in knot theory.
    }
    \label{fig:topological_characterization}
\end{figure}

Topological characterization of a chaotic attractor using Unstable Periodic Orbits (UPOs) has been performed since decades now, especially for the Rössler system \cite{Ross76} for which it has been done in 1995 \cite{Letellier_1995b}.
We complement this method to ensure that it is fully reproducible and applied it to the standard Rössler attractor \cite{rosalie2014toward, rosalie2016template}: Fig~\ref{fig:topological_characterization} details the steps to fully characterize the topological structure of a chaotic attractor using its topological invariants.

First, the flow of the attractor (Fig.~\ref{fig:topological_characterization}a) has to be displayed using a clockwise progression where a Poincaré section can be chosen accordingly to the bounding torus of the attractor (Fig.~\ref{fig:topological_characterization}b).
From this Poincaré section, a variable is chosen to build a first return map to the Poincaré section, this map (Fig.~\ref{fig:topological_characterization}c) is the discrete representation of the flow.
From this map, we are able to numerically extract unstable periodic orbits that have been followed during the numerical resolution of the system using Runge-Kutta 4 algorithm.
As these orbits are unstable, the numerical solution is closer enough to them that we can numerically extract their trajectory in the phase space (Fig.~\ref{fig:topological_characterization}d) using starting points from the first return map where they are labeled with symbols according to the branches with 0 or 1 for each passage through the Poincaré section (Fig.~\ref{fig:topological_characterization}c).
These orbits are considered as the central element of the attractor and the purpose of this method is to characterize the solution by finding a template that permits to compute their linking numbers theoretically (Fig.~\ref{fig:topological_characterization}e). 
This template is validated by ensuring that numerical and theoretical linking numbers between each pair of orbits are the same.
The template of this attractor is made of one stretching and folding mechanism (one of the simplest chaotic mechanism).
The linking matrix provides a numerical description of the template representation where the link between period--1 and period--2 orbits are illustrated (Fig.~\ref{fig:topological_characterization}e); this matrix details the permutation and torsion occurring to the flow of the attractor from one passage through the Poincaré section to another.
This method has been thoroughly tested and works perfectly for strongly dissipative systems. For other systems that do not share this property, such as the Lorenz--84 attractor, the use of UPOs remains a topological invariant that ensures a valid template regardless of the method used to obtain it.

\subsection{Color tracer mapping}

First return maps are well adapted to characterize strongly dissipative systems for which a flat or almost flat Poincaré sections can be obtained.
In such conditions, a single scalar is sufficient to characterize a trajectory crossing the Poincaré section, and a first return map can be used to study the successive returns.
It is no longer the case when the system is weakly dissipative because the Poincaré section becomes bidimensional which makes it difficult to reduce the structure to a classical first return map.

To account for the bidimensional dimension of the Poincaré section, colored tracer can be used as a color map by tracing the flow between successive Poincaré sections \cite{MangiaroHDR2014}.
To do so, let us define the two-dimensional map
\begin{equation}
    \begin{array}{l}
    g \colon \mathbb{R}^n ~~ \to ~~ \mathbb{R}^n \\
    ~~~~~~ \vec{X}_{j} = \begin{pmatrix} x_j \\ y_j \\ \vdots \end{pmatrix} ~~ \mapsto ~~ \vec{X}_{j+1} = \begin{pmatrix} f_x(\vec{X}_{j}) \\ f_y(\vec{X}_{j}) \\ \vdots \end{pmatrix} \;,
    \end{array}
    \label{eq:map}
\end{equation}
with $j \in \mathbb{N}$ the incremental time,
and
\begin{equation}
    \begin{array}{l}
    C^{\alpha} \colon \mathbb{R}^n ~~ \to ~~ \mathbb{R}^3 \\
    ~~~~~~~~ \vec{X} = \begin{pmatrix} x \\ y \\ \vdots \end{pmatrix} ~~ \mapsto ~~ C^{\alpha}(\vec{X}) = \begin{pmatrix} f_R^{\alpha}(\vec{X}) \\ f_G^{\alpha}(\vec{X}) \\ f_B^{\alpha}(\vec{X}) \end{pmatrix} \;,
    \end{array}
    \label{eq:col}
\end{equation}
the $\alpha$-color palette with $f_R^{\alpha}$, $f_G^{\alpha}$ and $f_B^{\alpha}$, respectively the red, green and blue functions used to associate colors with the phase space.
Then, the color map $\Gamma_p^\alpha$ can be defined as
\begin{equation}
    \begin{array}{l}
    \Gamma_p^{\alpha} \colon \mathbb{R}^{n+3} ~~ \to ~~ \mathbb{R}^{n+3} \\
    ~~~~~~~~ \{\vec{X}_j, C^{\alpha}(\vec{X}_j) \} ~~ \mapsto ~~ \{\vec{X}_{j+p}, C^{\alpha}(\vec{X}_j) \} \;,
    \end{array}
    \label{eq:colmap}
\end{equation}
where $p \in \mathbb{N}$ denotes the $p^{th}$ return.

In practice, it is not necessarily required to define the map $g$ explicitly when it can be reconstructed from the Poincaré map of a given flow as it will be the case.
Given a set of points $\Sigma_0$ of a chosen Poincaré section $P$ defined as $\Sigma_0 = \mathcal{A} \cap P$ with $\mathcal{A}$ the studied attractor, each point of this section at coordinates $(x_j,y_j,z_j)$, will come back to the section at the next iteration with new coordinates $(x_{j+1},y_{j+1},z_{j+1})$, leading to a reorganization $\Sigma_1$ of the original set of points.

To reconstruct the attractor structure, the method then consists in applying the color map from two successive iterations $i$ and $i+1$ and then to reconstruct the color evolution manually as a continuous transformation from $\Gamma_j$ to $\Gamma_{j+1}$.
Since the tracer is invariant in time (only the coordinates of the color tracer will change between two iterations), it can be used both back and forth.

In its principle, this technique can be applied with any color palette $\alpha$. For the same reason, it can also be considered between any successive iterations, with palettes defined either locally or globally.
Various coloring descriptions can be used to analyze and validate the reconstruction.

\section{The Lorenz--84 attractor}
\label{sec:lorenz_84_attractor}

\begin{figure}[htbp]
    \centering
    \includegraphics[width=.35\textwidth]{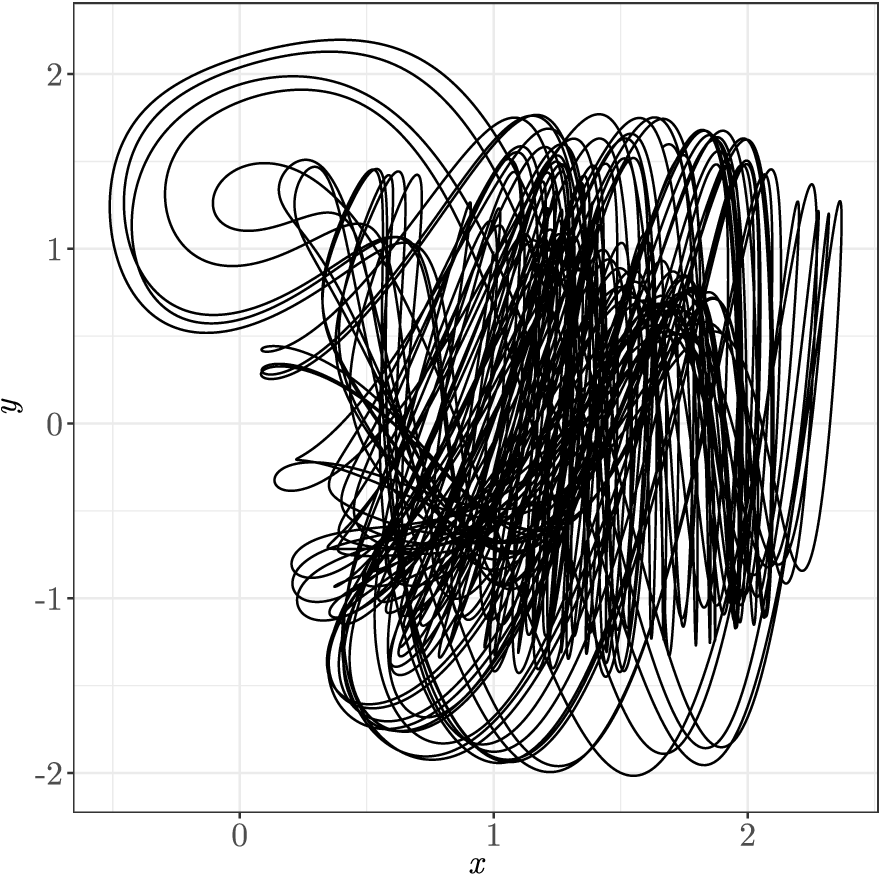} \quad
    
    (a) $(x-y)$ phase space of the Lorenz--84 attractor.\medskip
    
    \includegraphics[width=.35\textwidth]{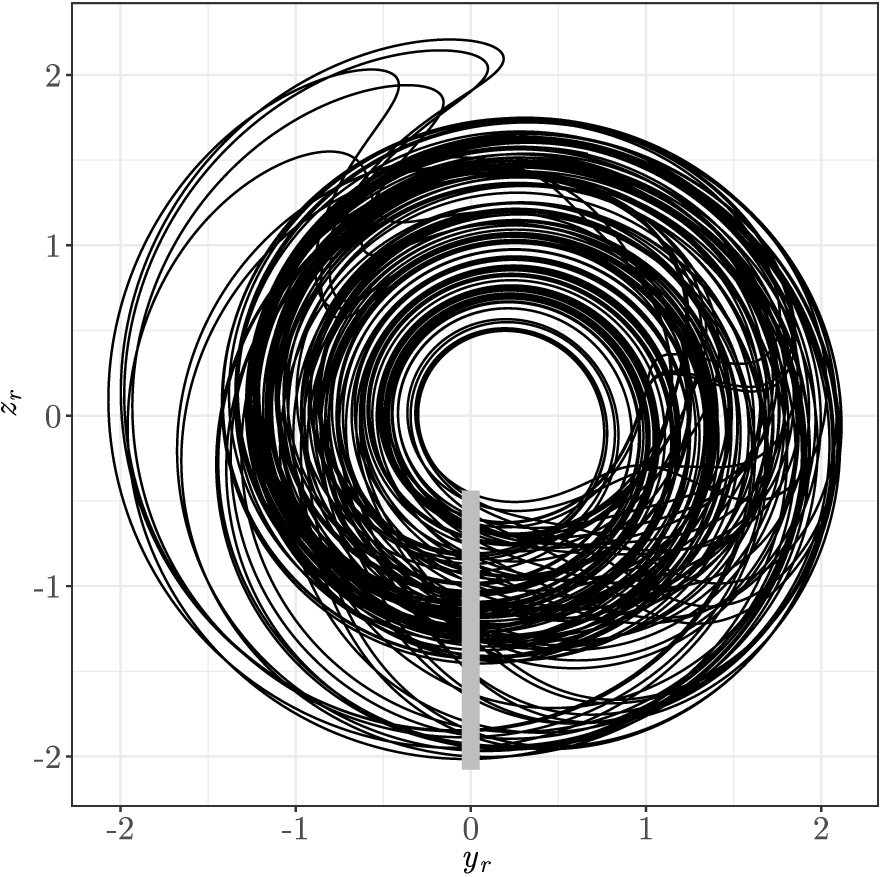}
    
    (b) Rotated attractor.\medskip
    
    \includegraphics[width=.35\textwidth]{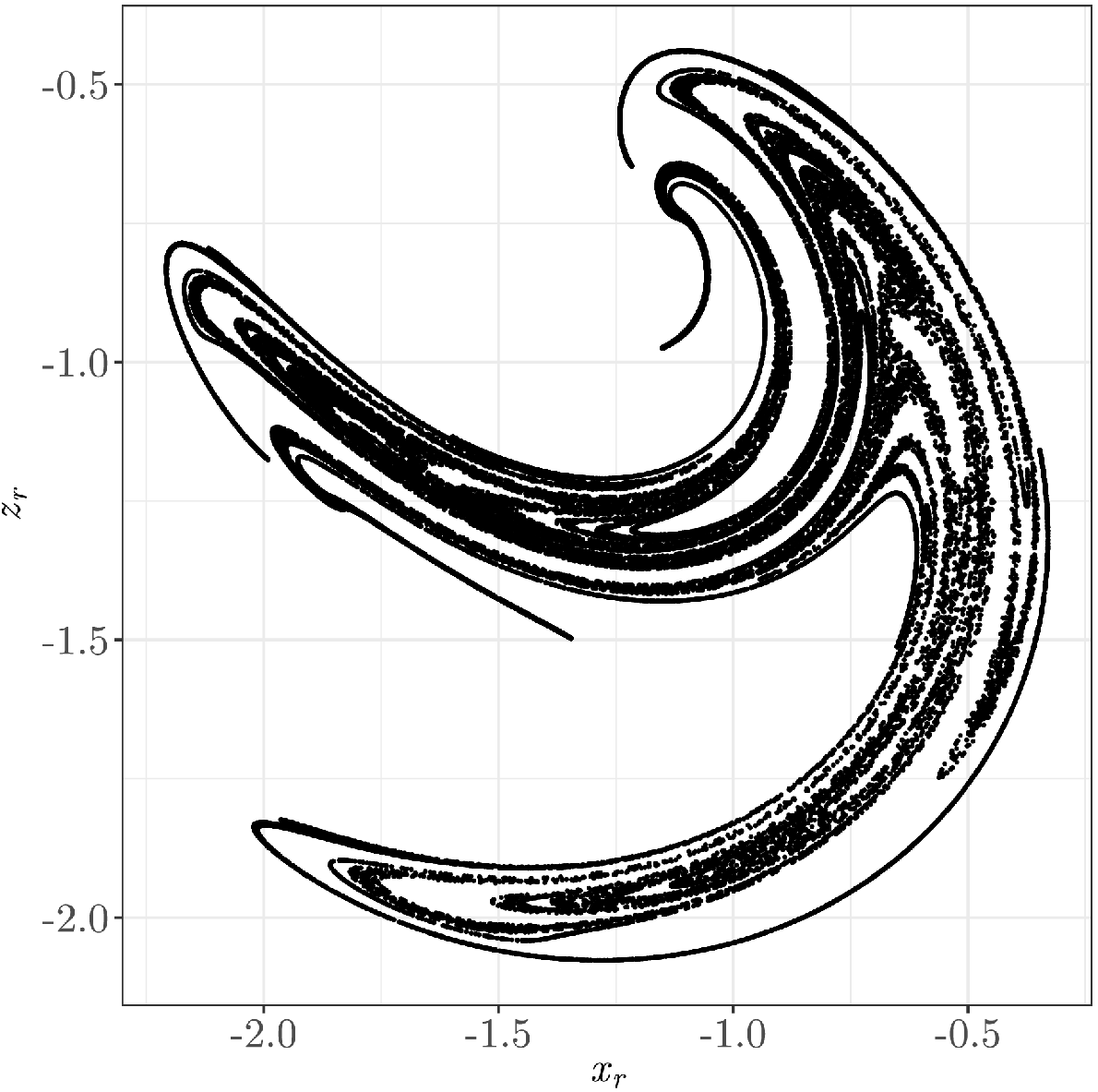}
    
    (c) Poincaré section
    \caption{
        Lorenz--84 attractor solution to system \eqref{eq:lorenz84} for the parameters $a=0.25$, $b=4.0$, $F=8.0$ and $G=1.0$.
        (a) $(x, y)$ projection of the attractor.
        (b) $\mathcal{L}$, the rotated Lorenz--84 attractor $(x_r, y_r, z_r)$ solution to system
        \eqref{eq:lorenz84} where the flow evolves clockwise in the projection plane ($y_r, z_r$).
        The grey line is the Poincar\'e section \eqref{eq:lorenz_02_section}.
        (c) Poincaré section.
    }
    \label{fig:lorenz84_01_attra}
\end{figure}

Introduced in 1984, the Lorenz--84 system is described as follows:
\begin{equation}
    \left\{
    \begin{array}{l}
    \dot{x} = - y^2 - z^2 -a x + aF \\
    \dot{y} = xy - bxz - y + G \\
    \dot{z} = bxy +xz -z \;.
    \end{array}
    \right.
    \label{eq:lorenz84}
\end{equation}
For the parameters $a=0.25$, $b=4.0$, $F=8.0$ and $G=1.0$, the solution to this system is represented Fig.~\ref{fig:lorenz84_01_attra}a.
The attractor is studied with respect to conventions \cite{rosalie2014toward}: clockwise flow and standard insertion, indicating that, in the template, the right branch is up to the left branch. 
These conventions ensure that the template obtained can be compared with others.
The first step consists in rotating the attractor around the $y$-axis  $(x', y', z') = (-x, y, -z)$ that is combined with another rotation of angle $\theta = 1.5$ radian in phase plane $(y', z')$. This leads to the following variables:
\begin{equation}
    \left\{
    \begin{aligned}
    x_r &= -x \\
    y_r &= y \cos \theta + z \sin \theta \\
    z_r &= y \sin \theta - z \cos \theta
    \end{aligned}
    \right.
\end{equation}

Thus, $\mathcal{L}$ is an attractor where the flow evolves clockwise (Fig.~\ref{fig:lorenz84_01_attra}b). 

We start performing the topological characterization of the attractor $\mathcal{L}$ by first defining a Poincaré section:
\begin{equation}
    \mathcal{P} = \{ (x_{r,n}, z_{r,n}) ~ | ~ y_{r,n} = 0, ~\dot{y}_{r} < 0\} \;.
    \label{eq:lorenz_02_section}
\end{equation}
This section is displayed in Fig.~\ref{fig:lorenz84_01_attra}c where the weakly dissipative property of the attractor is clearly underlined by the bidimensional distribution.
For points in the Poincaré section, we thus build a variable $\rho_n$ from variable $z_r$ in a way that if $z_r$ corresponds to a value close to the center, then $\rho_n$ is closer to $0$ and reciprocally, when $z_r$ is close to the external boundary of the attractor then $\rho_n$ is close to $1$:
\begin{equation}
\label{eq:rho_n}
\rho_n = \frac{-z_r-\min(z_r)}{\max(z_r)-\min(z_r)}
\end{equation}

\begin{figure}[htbp]
    \centering
    \includegraphics[width=.37\textwidth]{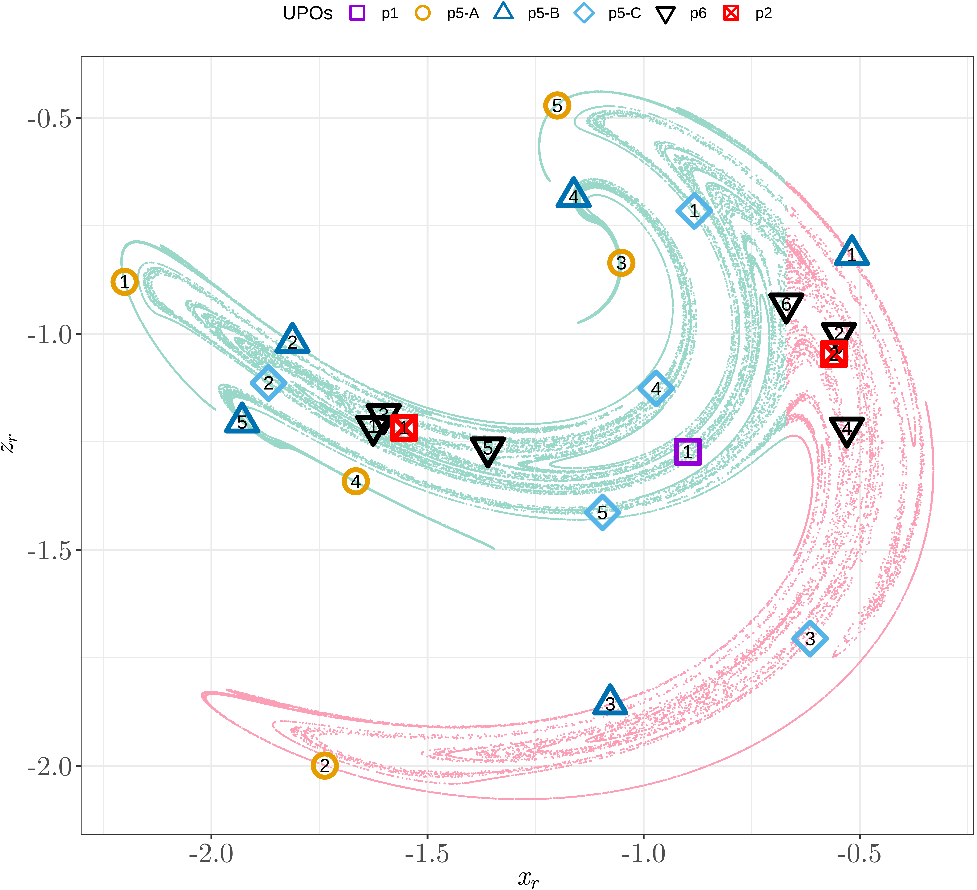}
    
    (a) Poincaré section with UPOs\medskip
    
    \includegraphics[width=.37\textwidth]{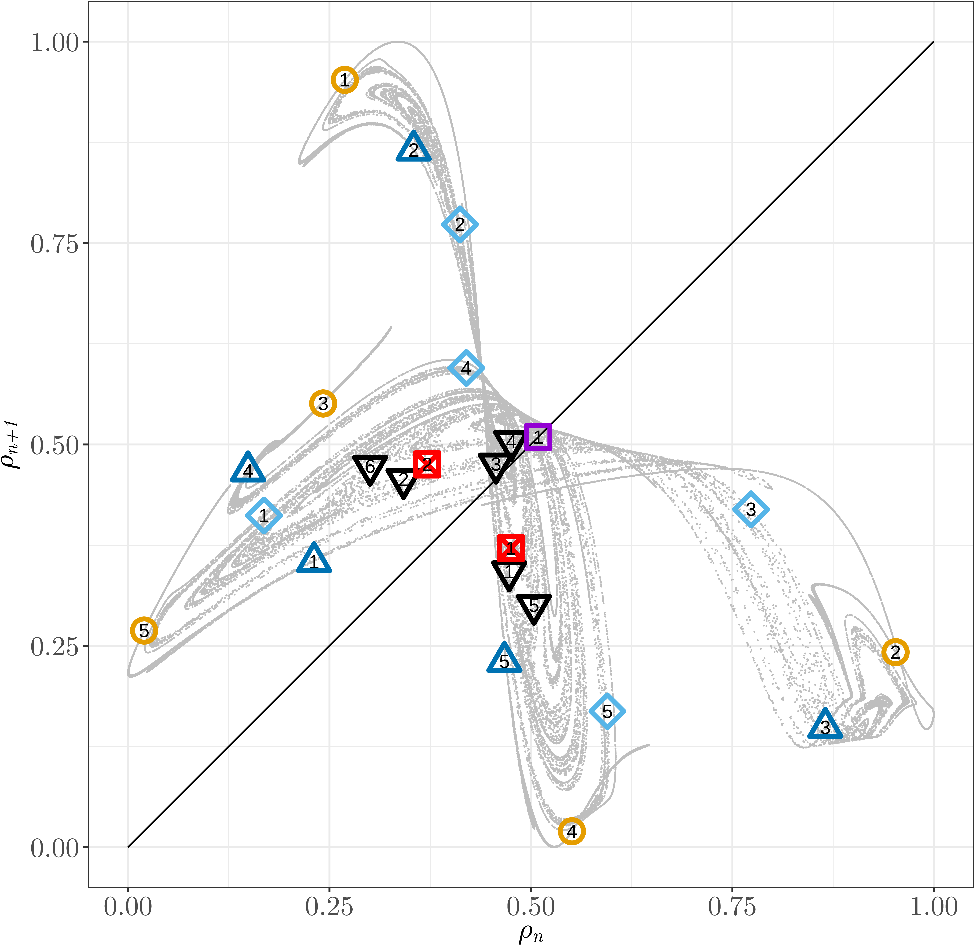}
    
    (b) First return map using $\rho_n$ \medskip
    
    \includegraphics[width=.37\textwidth]{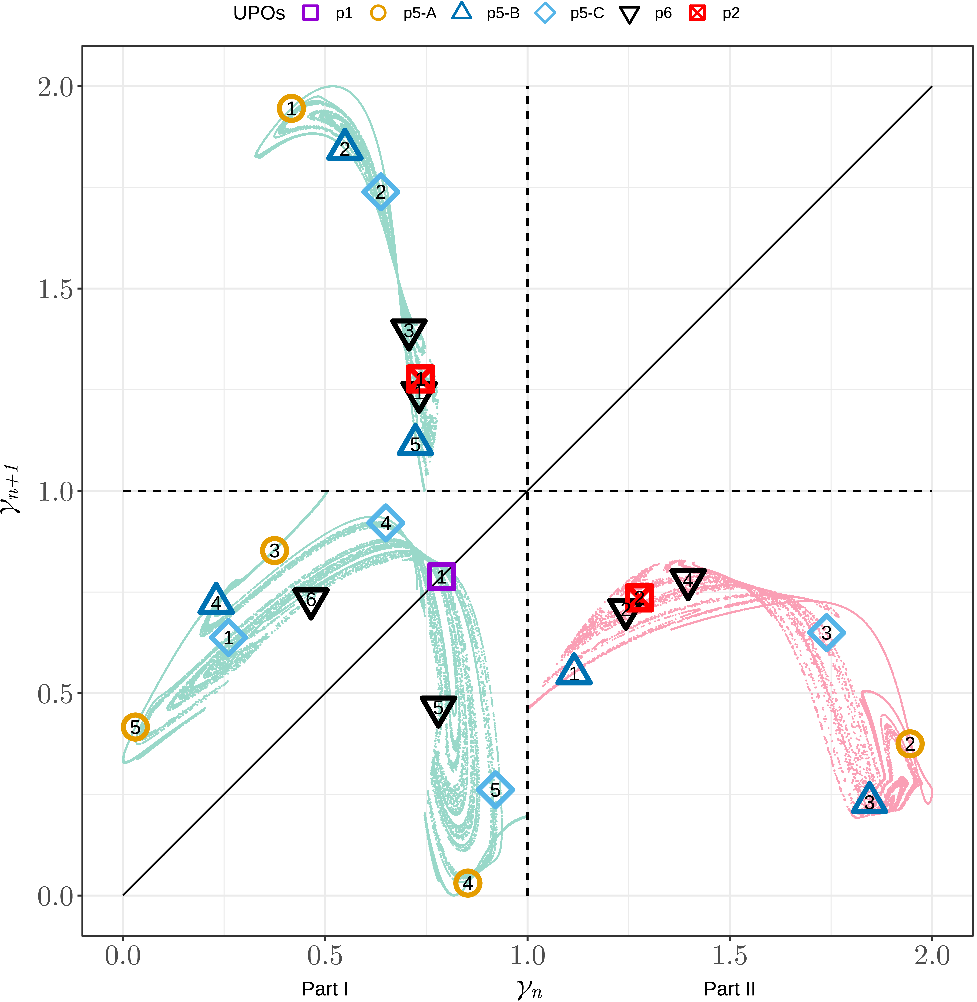}
    
    (c) First return map using $\gamma_n$
    
    \caption{
        (a) Poincaré section \eqref{eq:lorenz_02_section} of the attractor $\mathcal{L}$. 
        (b) First return map to the Poincaré section (a) using $\rho_n$.
        (c) First return map to the Poincaré section (a) using $\gamma_n$ \eqref{eq:gamma_n}.
        Part I (color green) is the first part of \eqref{eq:gamma_n} while part II (color pink) is the second line of \eqref{eq:gamma_n}.
        This colored partition is also visible in (c) clarifying the relative position of periodic points of (b).
    }
    \label{fig:lorenz84_02_section}
\end{figure}

A first return map can be plot using $\rho_n$ (Fig.~\ref{fig:lorenz84_02_section}a) as it is done for dissipative system (see Fig.~\ref{fig:topological_characterization} for the method).
However, the weakly dissipative property of $\mathcal{L}$ implies a loose of bijection property using $\rho_n$.
This impose us to extract periodic orbits directly from the Poincaré section instead of using a first return map.
Five periodic orbits are numerically extracted from the first return map: one period--1 orbit, three period--5 orbits and one period--6 orbit (Fig.~\ref{fig:lorenz84_02_section}a).
Using equation \eqref{eq:rho_n}, the periodic orbits are represented in a first return map (Fig.~\ref{fig:lorenz84_02_section}b).
The linking numbers between each couple of orbits are presented in Tab.~\ref{tab:lorenz84_02_lk}. 
This topological invariant indicates how the pairs of unstable periodic orbits are twisted together.
The first return map is much more complex given the properties of $\mathcal{L}$ when compared to the first return map of the Rössler system (Fig.~\ref{fig:topological_characterization}c).
Nevertheless, we observe thick increasing and decreasing branches, reflecting the dynamic structure of the attractor. 
The presence of this thickness in the branches complicates the interpretation of the map, especially since it appears that stretching occurs simultaneously in multiple directions. 
To better study the structure of $\mathcal{L}$, we chose to use two components to construct a variable representing the Poincaré section. 
This technique of Poincaré section made of multiple components is usually used when the bounding torus of the attractor has a genus higher than one leading to multiple cuts of the bounding torus to have a complete representation of the Poincaré section \cite{rosalie2014toward}.
It also applies here.
Similarly to $\rho_n$ (Eq. \ref{eq:rho_n}) we choose to construct $\gamma_n$ with the orientation from the inside of $\mathcal{L}$ to the outside boundary:
\begin{equation}
\label{eq:gamma_n}
\gamma_n = \left\{
\begin{aligned}
&\frac{-z_r-\min(z_r)}{1.5-\min(z_r)} & \text{ if } z_r \geq -1.5 \text{ and } x_r \leq -0.67 \\
&\frac{-z_r-\min(z_r)}{\max(z_r)-\min(z_r)} + 1 & \text{else}\\
\end{aligned}
\right.
\end{equation}
This partition of the Poincaré section is highlighted in Fig.~\ref{fig:lorenz84_02_section}a with two colors in the section that are also reported in Fig.~\ref{fig:lorenz84_02_section}c of the first return map using $\gamma_n$.
The latter map is made of four parts with values of $\gamma_n$ between 0 and 1 that correspond to the upper left part of the application (Fig.~\ref{fig:lorenz84_02_section}a) labeled as part I, while the rest of the map with values between 1 and 2 corresponds to part II.
The fact that there is no point in the upper right corner means that there is no consecutive points in the first return map from part II to itself.
Reciprocally, there are points in the bottom left corner, this indicates that there are transitions from part I to part II.
This first return map also indicates that if a trajectory goes to part II, its points will be in part I at the next return.
To conclude this first step of the topological characterization, it seems that almost six branches (two per area) can be distinguished, but the thickness of the first return map combined with the artificial partition made using multiple components for the Poincaré section prevent us from directly pursuing the analysis in the formal way (Fig.~\ref{fig:topological_characterization}).

\begin{table}[t]
    \centering
    \caption{
        Linking numbers between pairs of orbits
        extracted from the chaotic attractor $\mathcal{L}$.
        The period--2 orbits is not included in the attractor but located in the period--2 cavity inside the attractor (see Fig.~\ref{fig:P2} and Sec.~\ref{sec:bounding_torus_and_toroidal_chaos} for details).
    }
        \begin{tabular}{cccccc}
            \\[-0.3cm]
            \hline \hline
            \\[-0.3cm]
            & (p1) & (p5-A) & (p5-B) & (p5-C) & (p6) \\[0.1cm]
            \hline
            \\[-0.3cm]
            (p5-A)  & 2 & \\[0.1cm]
            (p5-B)  & 2 & 10 & \\[0.1cm]
            (p5-C)  & 2 & 10 & 10 \\[0.1cm]
            (p6)    & 3 & 12 & 12 & 12 \\[0.1cm]
            (p2)    & 1 & 4 & 4 & 4 & 5 \\[0.1cm]
            \hline \hline
        \end{tabular}
    \vspace{-.5em}
    \label{tab:lorenz84_02_lk}
\end{table}

\section{Skeleton analysis}

\begin{figure}[htbp]
    \centering
    \qquad \qquad
    \begin{tabular}{ccc}
    \includegraphics[width = .37\textwidth]{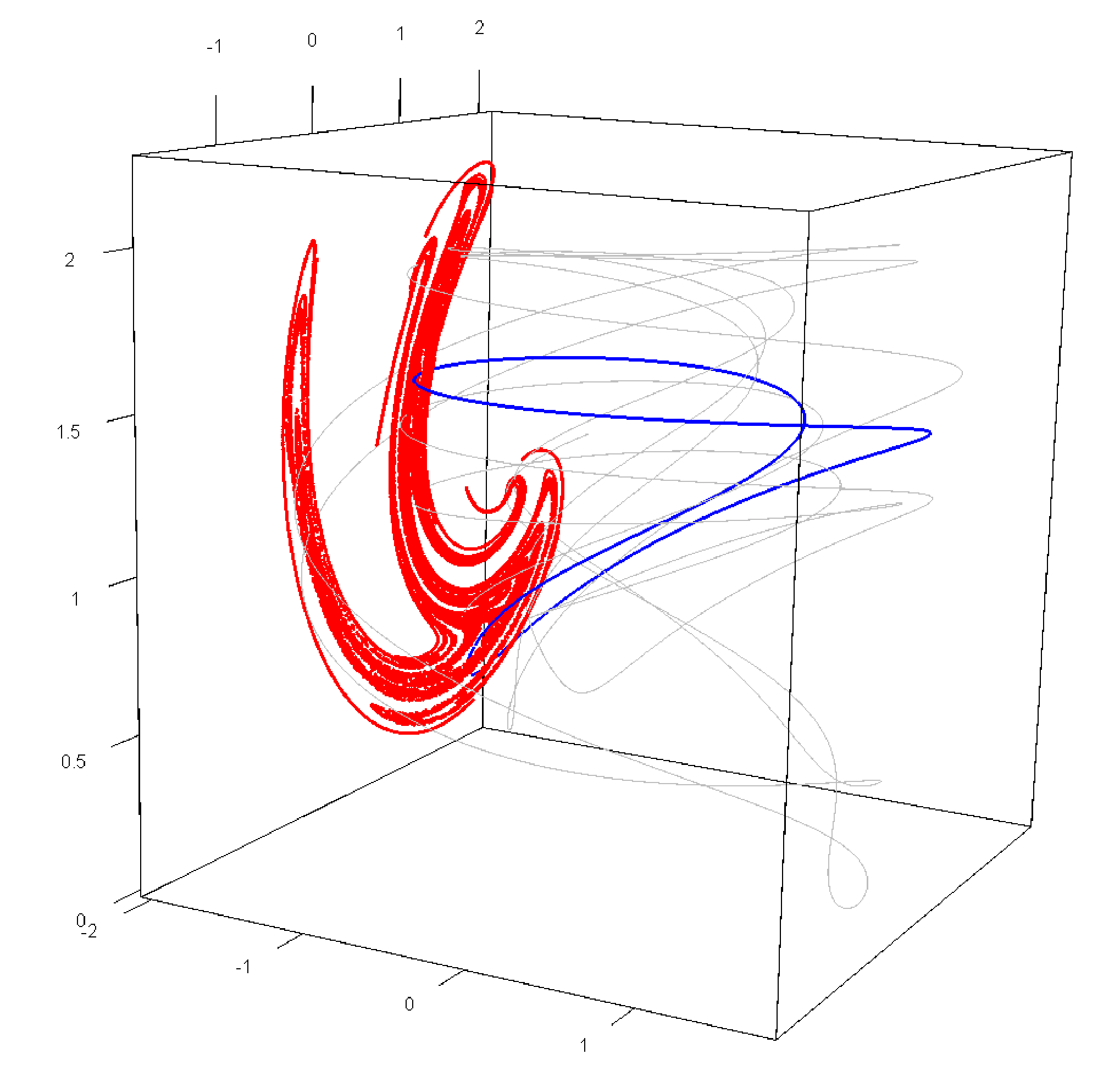} \\
    (a) Phase space \\
    \includegraphics[width = .46\textwidth]{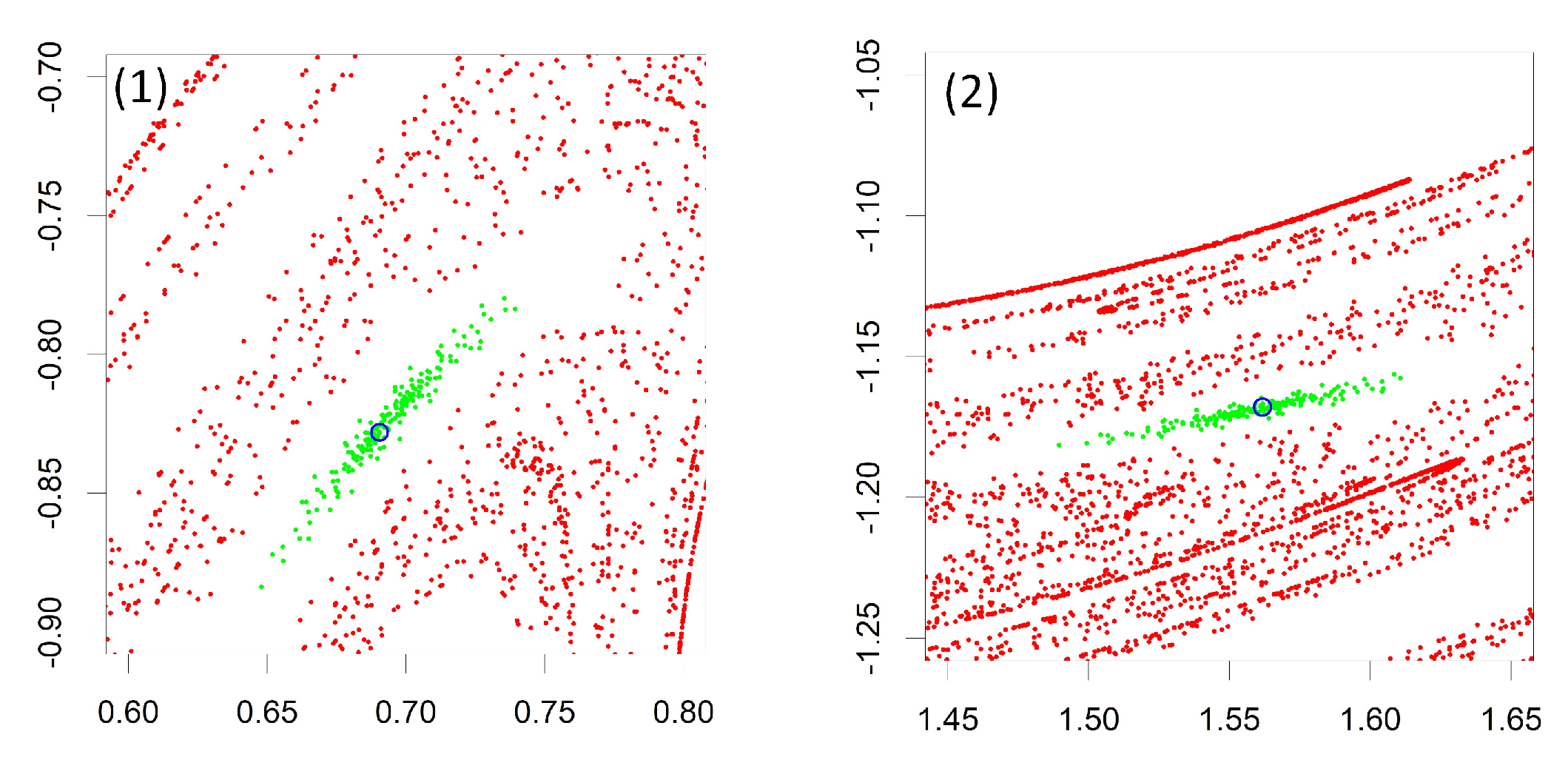} \\
      (b) Region around the period--2 orbit \\
    \end{tabular}
    \caption{
    The period--2 orbit (in blue) represented passing through a Poincaré section (in red) of the Lorenz--84 attractor (a) (a part of the Lorenz--84 attractor is also represented in light gray). Zooms on the Poincaré section are also presented (b) at the vicinity of the period--2 orbit crossings (blue circle). The crossing points (in green) around the period--2 orbit belong to the long transient to reach the Lorenz-84 attractor (in red).
    }
    \label{fig:P2}
    \vspace{-2em}
\end{figure}

\subsection{Bounding torus and toroidal chaos}
\label{sec:bounding_torus_and_toroidal_chaos}

One important feature of the period--6 orbits is their vicinity and their separation in two groups both organized around an apparently empty region of the Poincaré section, suggesting an organization around a period--2 orbit which was however not found when studying the first return map (see previous section).
To investigate this question, an ensemble of simulations starting from this empty area was run that all supported the vicinity of a period--2 orbit and the existence of cavity around it.
The period--2 orbit was then extracted using an optimization technique.
Its trajectory in the phase space is presented in Figure \ref{fig:P2}a together with the Poincaré section that shows the two crossings.
Zooms on the Poincaré section at the crossing points are also presented in Fig.~\ref{fig:P2}b and \ref{fig:P2}c).
To confirm the existence of the cavity around this period--2 orbit, a trajectory of long duration was run starting from the period--2 orbit.
A very long transient was required to get out of the hole, revealing a region of quasi-stability (361 return maps were necessary to escape from the hole and reach the attractor region).
Once reaching the Lorenz attractor, no trajectory could ever come back to this specific region.
These properties reveal that this region is part of the attraction basin of the Lorenz-84 attractor, is located inside the attractor (as a cavity), but is not part of the attractor.
From a topological point of view, this means that the Lorenz-84 attractor is organized around a toroidal cavity of period--2.

This proves that the Lorenz--84 is a toroidal attractor since it is bounded outside by a genus--1 torus clearly visible in Figure \ref{fig:lorenz84_01_attra}b, and inside by genus--1 cavity of period--2.
To the best of our knowledge,
it is the first case of toroidal chaos revealed to be organized around a cavity of period--2.
The attractor is also organized around an unstable period--1 orbit presented in Fig.~\ref{fig:lorenz84_02_section}.
However, the regions of this period--1 orbit appears quite dense.
It seems unlikely possible to bound the inside part of the attractor by a genus--1 torus of period--1.

\begin{figure*}[htb]
    \centering
    \begin{tabular}{ccc}
    \includegraphics[width=.245\textwidth]{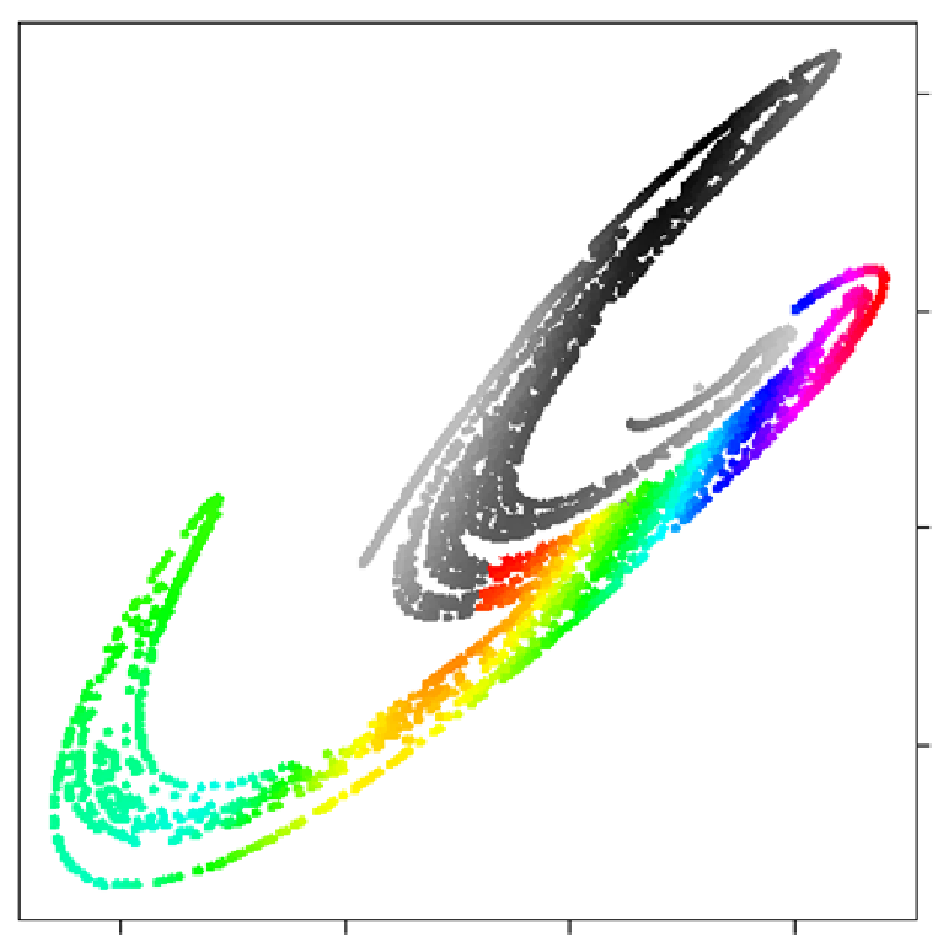} &
    \quad
    \includegraphics[width=.245\textwidth]{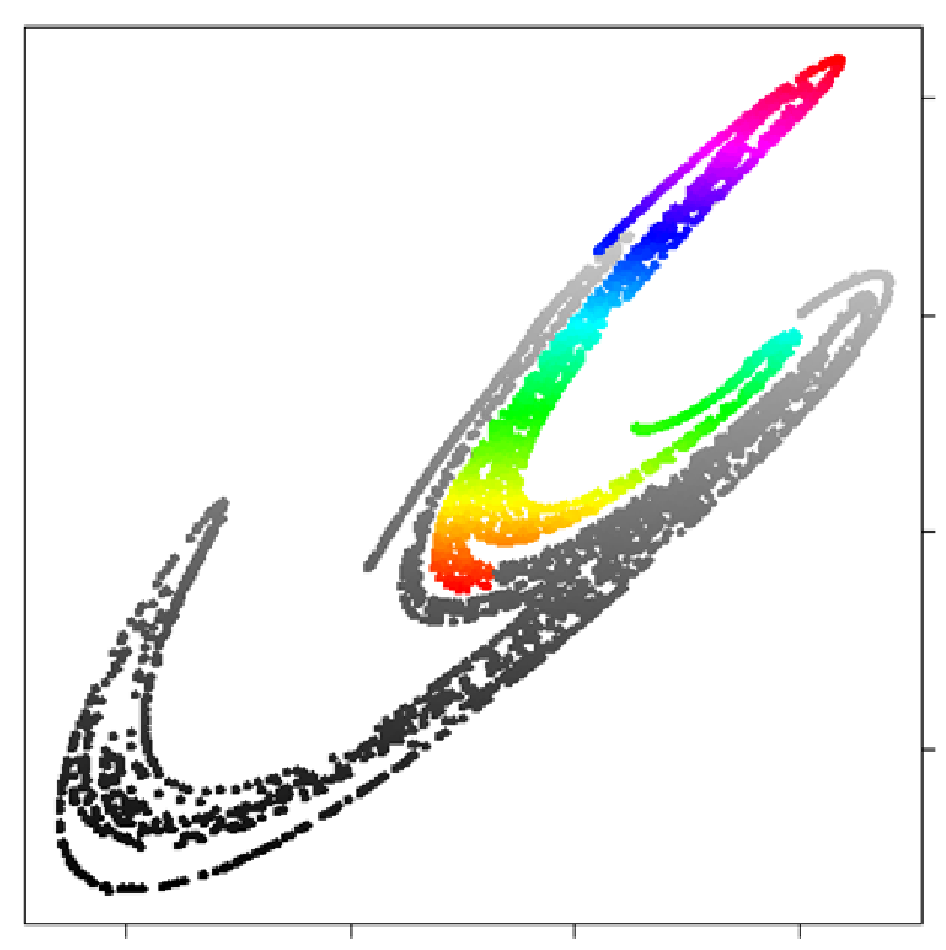} &
    \quad
    \includegraphics[width=.245\textwidth]{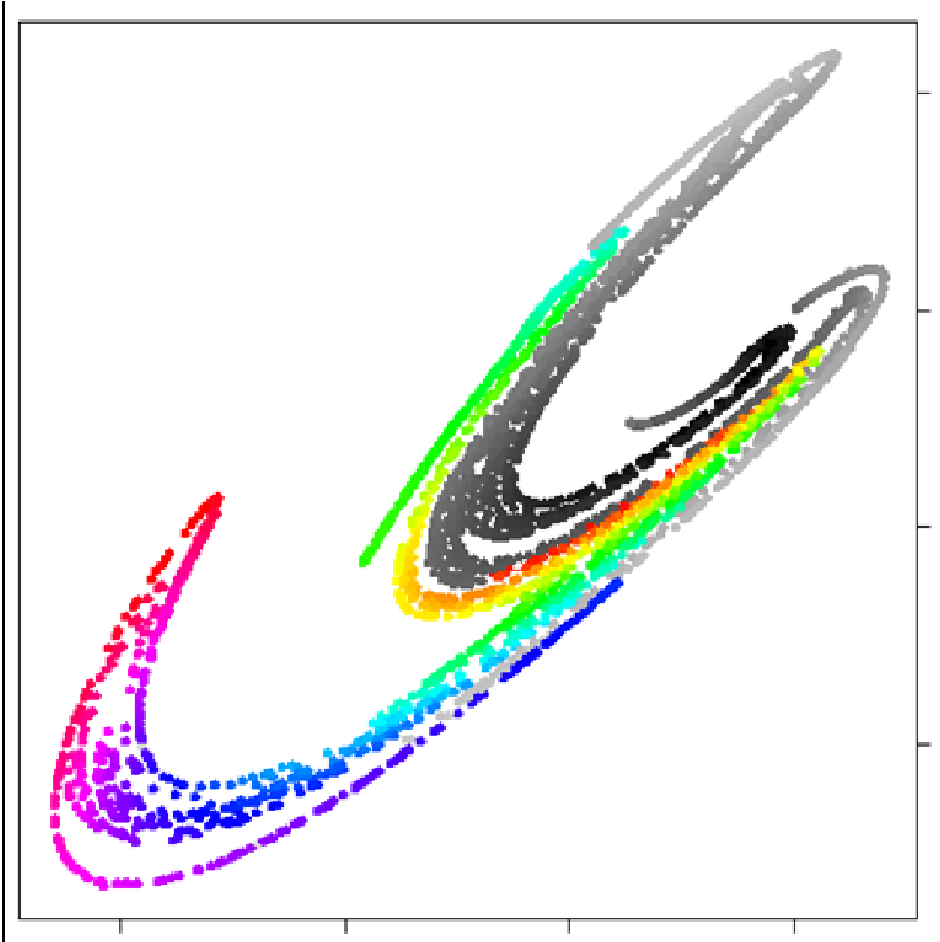} \\
      (a) $\Gamma^{~\alpha}_{n-1}$ &
      (b) $\Gamma^{~\alpha}_{n}$ &
      (c) $\Gamma^{~\alpha}_{n+1}$ \\
    \end{tabular}
    \caption{
Colored map $\Gamma^{\alpha}$ at three successive iterations using the $\alpha$-palette defined at iteration $n$ as a vertical gradient of colors in a restricted part of the section. Note that a $\beta$-palette is simultaneously used at iteration $n$ with a vertical gradient of gray, in the remaining part of the section.
    }
    \label{fig:ColoredLrz}
\end{figure*}

\begin{figure*}
    \includegraphics[width = .33\textwidth]{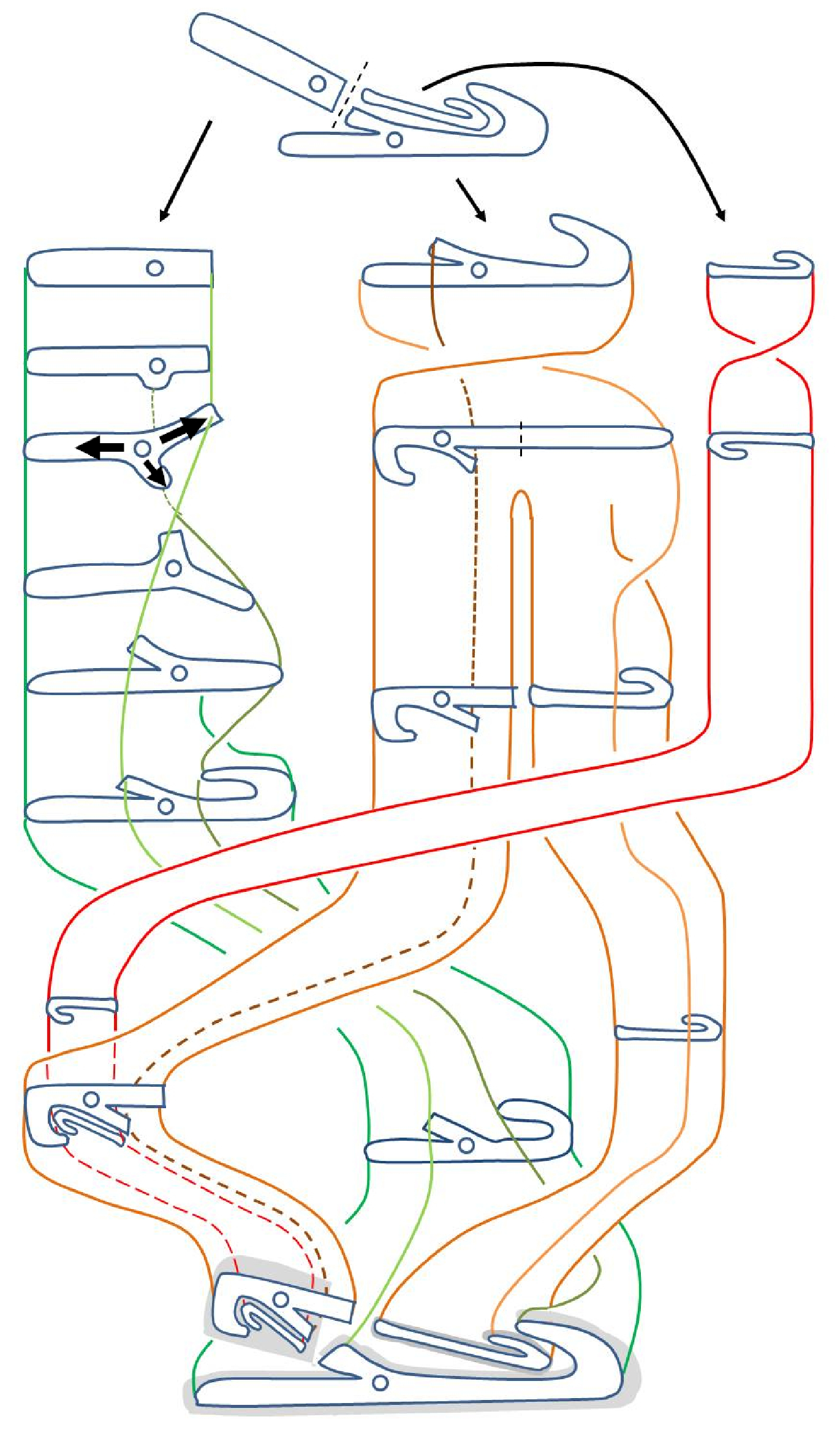}
    \put(-30,5){\includegraphics[width = .08\textwidth]{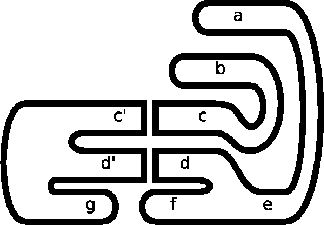}}
    \hspace*{1.1cm}
    \includegraphics[width = .32\textwidth, trim = {0 10 0 10}, clip]{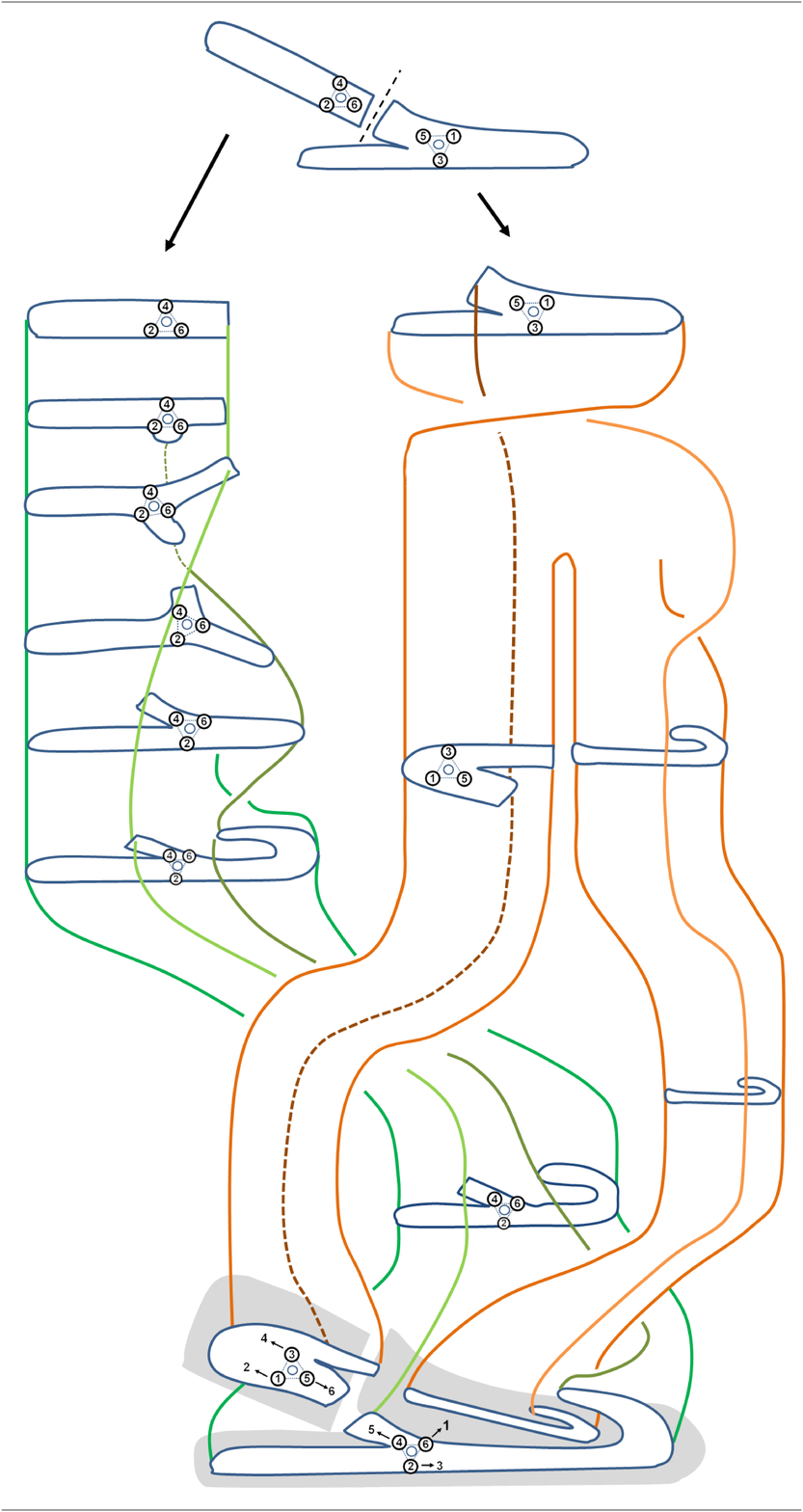}
    \hspace*{0.1cm}
    \includegraphics[width = .23\textwidth]{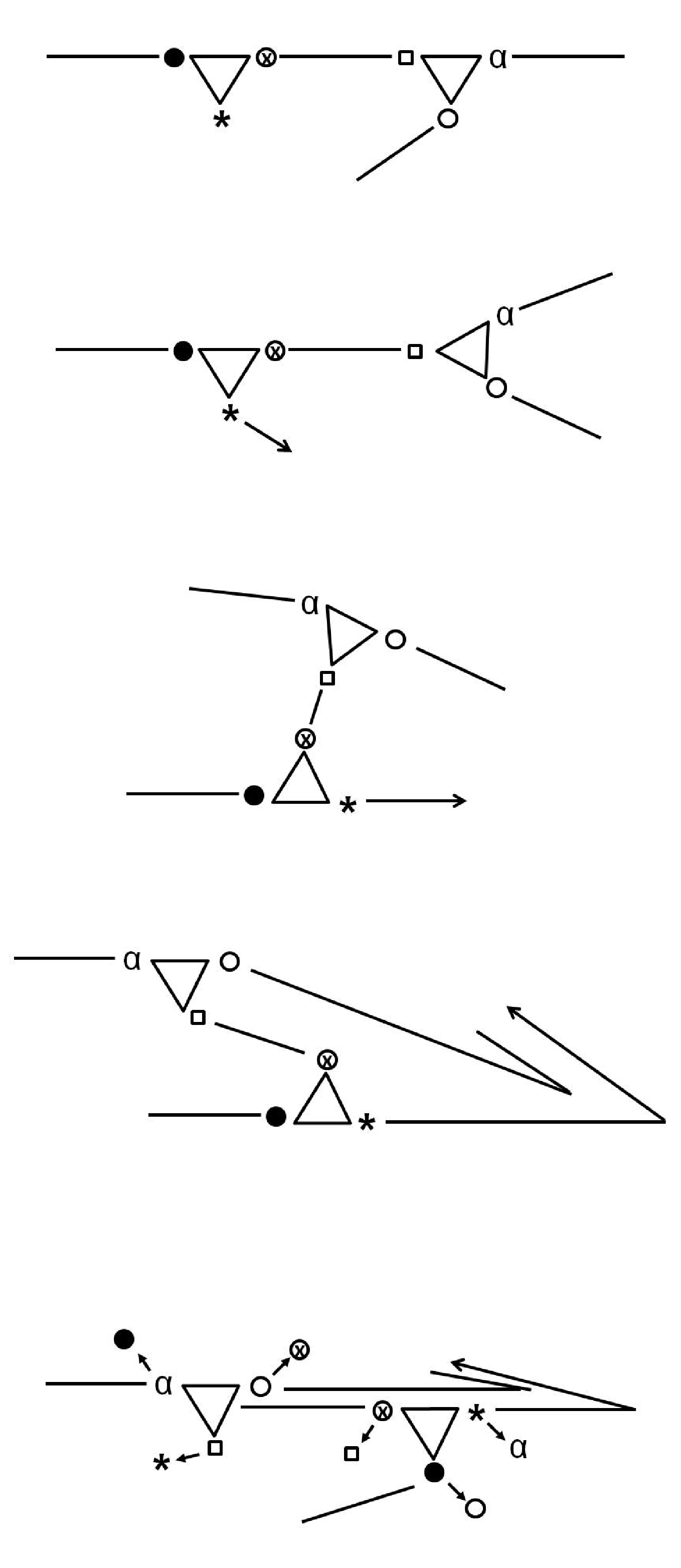} \\
        (a) Reconstructed skeleton \hspace*{5cm} \medskip \medskip
        (b) Simplified skeleton \hspace*{2cm} 
        (c) Flattened skeleton
    \caption{
    Skeleton deduced from the colored map analysis (a) with its multiple stretching represented by three black arrows. A simplified skeleton (very short branches $a$, $b$ and $g$ are removed since not required) with the period--6 unstable periodic orbit represented by triangles in and the period--2 cavity by circles in between (b). And a flattened version of the simplified skeleton (c).
    }
    \label{fig:skeleton}
\end{figure*}

\subsection{Colored tracer analysis}
\label{sec:colored_tracer}

The color tracer technique was then used to investigate the structure of the attractor more in depth.
In the present case, the color tracer was applied on a localized part of the Poincaré section as a vertical color gradient (Fig. \ref{fig:ColoredLrz}b) and then traced back (\ref{fig:ColoredLrz}a) and forth (\ref{fig:ColoredLrz}c).
Based on these three stages, the Poincaré section is completely covered by the colored tracers, including a partial overlap.
The attractor structure was then investigated by reconstructing manually the continuous transformation enabling to move from successive iterations. The resulting transformation is presented as a skeleton in Figure \ref{fig:skeleton}a.
No tearing was identified (since the color gradient remains continuous by increasing iterations). As a consequence, without any other information\cite{MangiaroHDR2014}, the topology can be reconstructed modulo 2$\pi$.
This ambiguity can be released here considering that the attractor is organized around a period--2 orbit characterized by a single positive twist.

For this first draft, nine branches are used to reconstruct the transformation, as shown in Fig.~\ref{fig:skeleton}a.
Although their limits may be unclear, the distinction between branches $c$-$c'$ and $d$-$d'$ is useful at this stage because the mixing will take place with distinct branches.
The analysis of the color map also confirms the binary behavior between the two sides of the attractor (left-right, right-left, left-right, etc.) resulting from the organization around the cavity of period--2 represented as circle in Figure \ref{fig:skeleton}a.

One very specific characteristic of this structure comes from the multidirectional stretching taking place in the upper left part of the section.
Such a multidirectional stretching mechanism was already reported in a preliminary study of the Lorenz--84 attractor \cite{MangiaroHDR2014,Mangiarotti_2016,Rosalie2023}, for which it is therefore a confirmation. Such a mechanism was also observed in the structure of the cereal crops attractor although associated with a tearing along one of the stretching direction in this latter case \cite{MangiaroHDR2014,Mangiarotti_2016}.
Since such a mechanism requires some thickness, it is very likely to be a common feature of weakly dissipative chaos and we do even conjecture that it is a necessary condition for it.
Note that multidirectional types of stretching were already discovered in higher-dimensional chaos \cite{Sciamarella1999,Sciamarella2001,Charo2022}, but this was less surprising in a space enabling several positive Lyapunov exponents.

A more in depth analysis reveals that this structure can be simplified.
Indeed, being weakly dissipative, a folding may remain visible after several cycles in the attractor.
However, each folding should be considered once.
Considering this simplification,
the reconstructed skeleton can be simplified as presented in Fig.~\ref{fig:skeleton}b.

Despite its thick structure, the possibility to represent the attractor's structure as flat branches was attempted using a succession sections assuming that the branches were flat (see Fig. \ref{fig:skeleton}c).
This representation is interesting because it clearly reveals three foldings, including one in the center.

\section{Theoretical templates}

\subsection{Squeezed template}
\label{sec:squeezed_template}

\begin{figure}[bt]
    \centering
    \includegraphics[width = .1\textwidth]{struct.eps}
    
    \includegraphics[width = .45\textwidth]{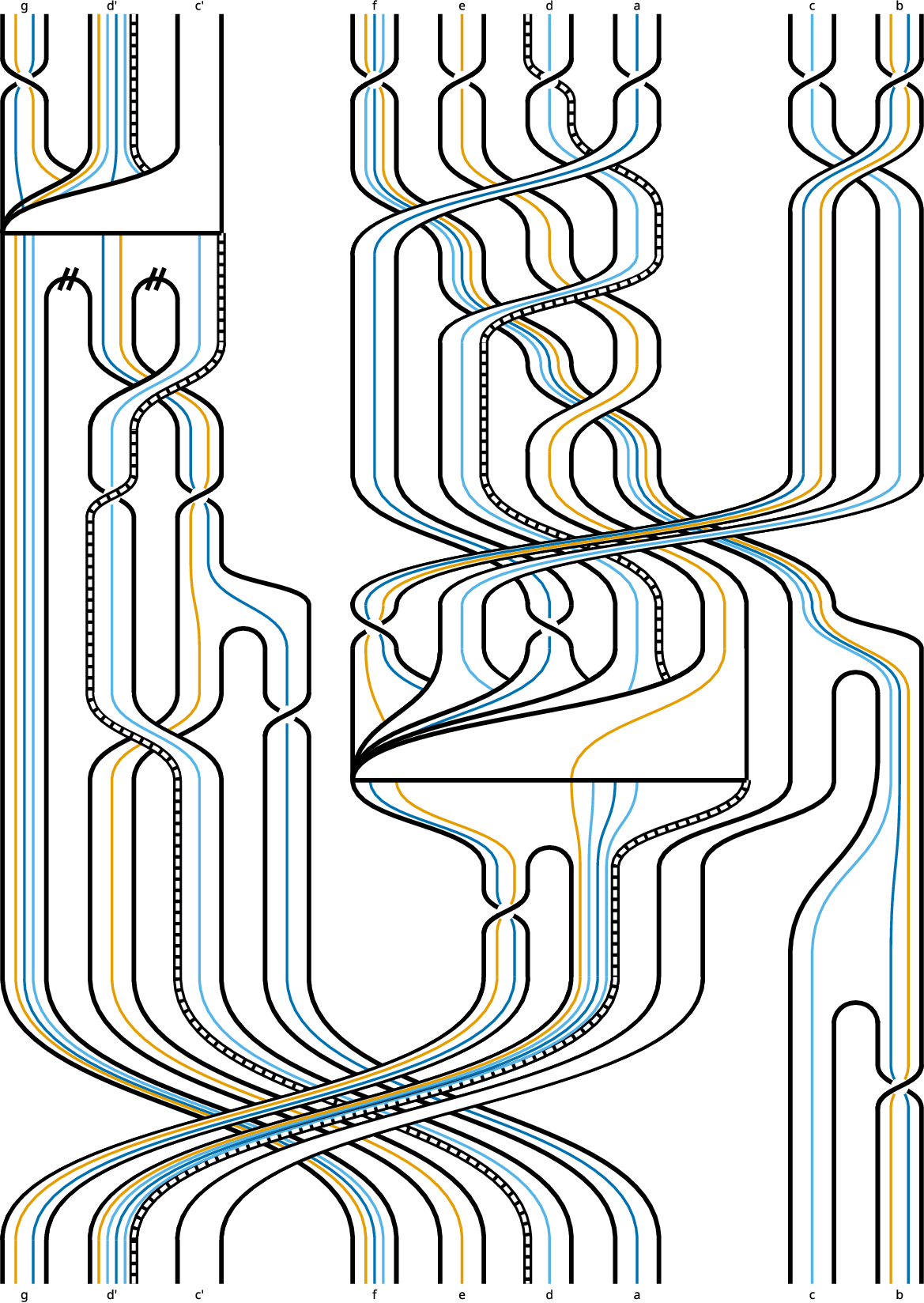}
    
    \includegraphics[width = .1\textwidth]{struct.eps}
    \caption{
    Template of the Lorenz--84 chaotic attractor with the period--5 unstable periodic orbits listed in Tab.~\ref{tab:lorenz84_02_lk}. The dashed line is the period--1 orbit.
    The color of the period--5 orbits are the colors used in Fig.~\ref{fig:lorenz84_02_section}.
    }
    \label{fig:template}
\end{figure}

Using the first draft obtained with the colored tracer method, a template is proposed.
Contrary to classical templates of dissipative attractors, this template (Fig.~\ref{fig:template}) details the transition from a structured partition of the Poincaré section to itself (see right side of Fig.~\ref{fig:template}).
The template contains seven branches ($a$, $b$, $c$, $d$, $e$, $f$ and $g$) including two branches split into two parts: $c$--$c'$ and $d$--$d'$. 
This enables us to create a plane version of the template that respects the Melvin and Tufillaro convention \cite{Melvin_1991} where the order from the left to the right corresponds to the order from the bottom to the top.
Moreover, the dashed line not only corresponds to the partition of branch $d$ and $d'$ but also to the period--1 orbit.
This partition is achieved in accordance with the usual chaotic mechanisms of tearing and stretching, and it also includes a multiple-direction stretching.
To underline this particular property in the template, we propose to add the symbol of parallelism to the folding mechanism to underline the fact that the stretching and folding occur simultaneously into two distinct directions.
Thus, from a single branch, this multiple-folding mechanism generates three separate branches by stretching them around a central cavity, requiring two separated holes in the template representation.
We validate this template using the linking numbers between pairs of orbits of periods 1 and 5 (Tab.~\ref{tab:lorenz84_02_lk}) because this template permits to predict theoretically the linking number obtained numerically.

This template is close to the template drawn from color tracer method. It contains several chaotic mechanisms: one multidirectional stretching mechanism on the left side, two folding mechanism in the middle and right side.
It also contains two stretching and squeezing mechanism on the left side and in the middle that could be concatenated \cite{Gilmore2016}.
For instance, the first hole that split the left and middle part of the template (at the top between $c'$ and $f$ which ends at the bottom between $a$ and $c$) can be simplified as a double positive twist using Reidemeister move.
In addition, the positive torsion can compensate a negative torsion occurring at the beginning of the template for branch $g$. 
These transformations help us to propose a reduced template in the continuation of this work.

\subsection{Reduced template}

\begin{figure}
    \centering
    \includegraphics[width = .45\textwidth]{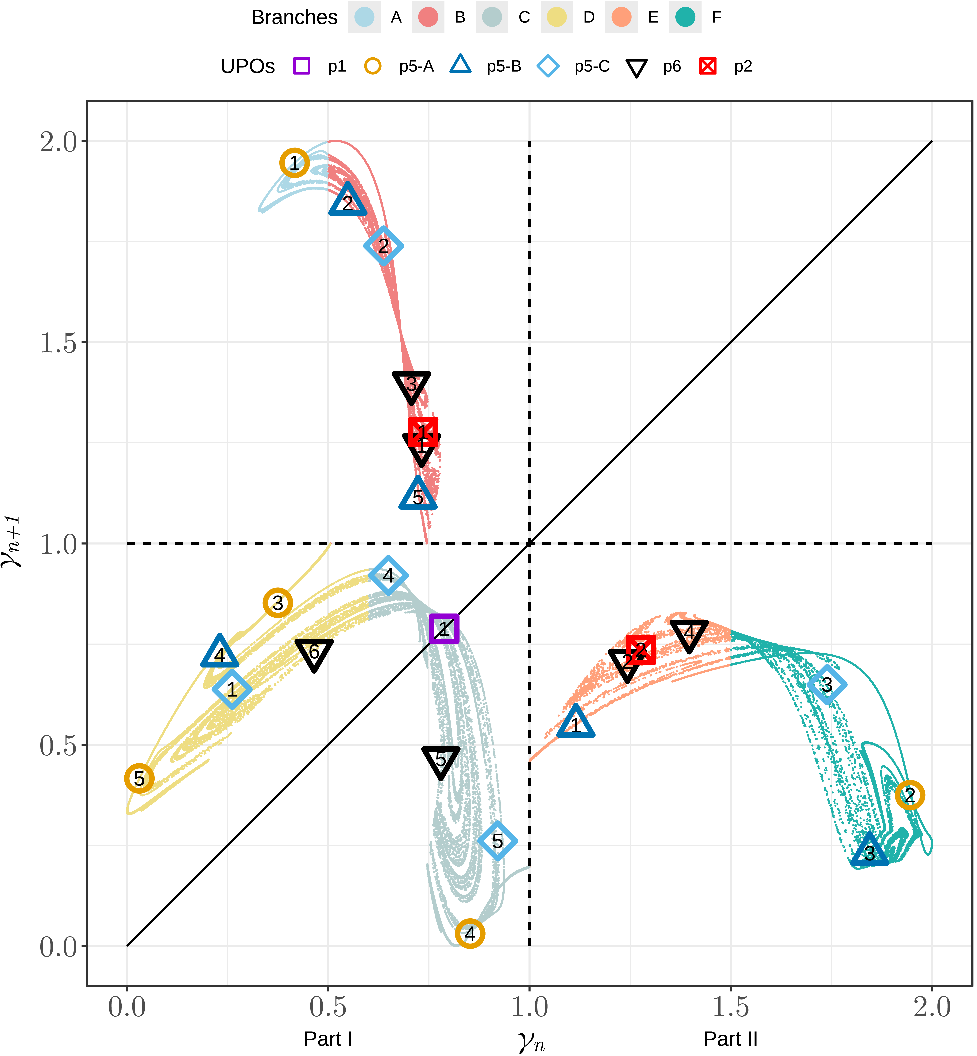}
    \caption{First return map using $\gamma_n$ of the Poincaré section \eqref{eq:lorenz_02_section}. Six branches are identified: A and B from part I to part II; D and C from part I to part I and, E and F from part II to part I. The encoding of the unstable periodic orbits are detailed in Tab.~\ref{tab:encoding} where the number in column code corresponds to the value in the symbol.}
    \label{fig:appli_6branches}
\end{figure}

The purpose of the reduced template is to validate a template by describing a synthesis of topological transformations occurring in the flow between two consecutive passages through the Poincaré section (see \cite{Letellier_2012} for dissipative systems).
The first return map to $\gamma_n$ (Eq. \ref{eq:gamma_n}) comes from a partition in two main parts (I and II) of the Poincaré section defined in Eq. \eqref{eq:lorenz_02_section}.
The first return map (Fig.~\ref{fig:appli_6branches}) shows that the points belonging to the decreasing branches $B$ and $C$ are split into two directions.
First the unimodal map labeled with symbols $A$ and $B$ indicates that there is a stretching and folding mechanism occurring from part I to part II.
Simultaneously, another unimodal map with $C$ and $D$ indicate another stretching and folding mechanism from part I to itself.
The flow continuously stretches and folds the trajectories into two directions from part I to either part I or part II.
Thus, the multidirectional stretching and folding mechanism that has been already identified is also visible in this first return map (Fig.~\ref{fig:appli_6branches}).
We can also deduce from it that the part II is also stretched and folded due to its unimodal mapping.
This first return map remains coherent while we have made this partition since the continuity is maintained for branches $B$ and $C$ as well as between branches $D$ and $E$ in addition to the unimodal shape.

\begin{figure}
    \centering
    \includegraphics[width = .45\textwidth]{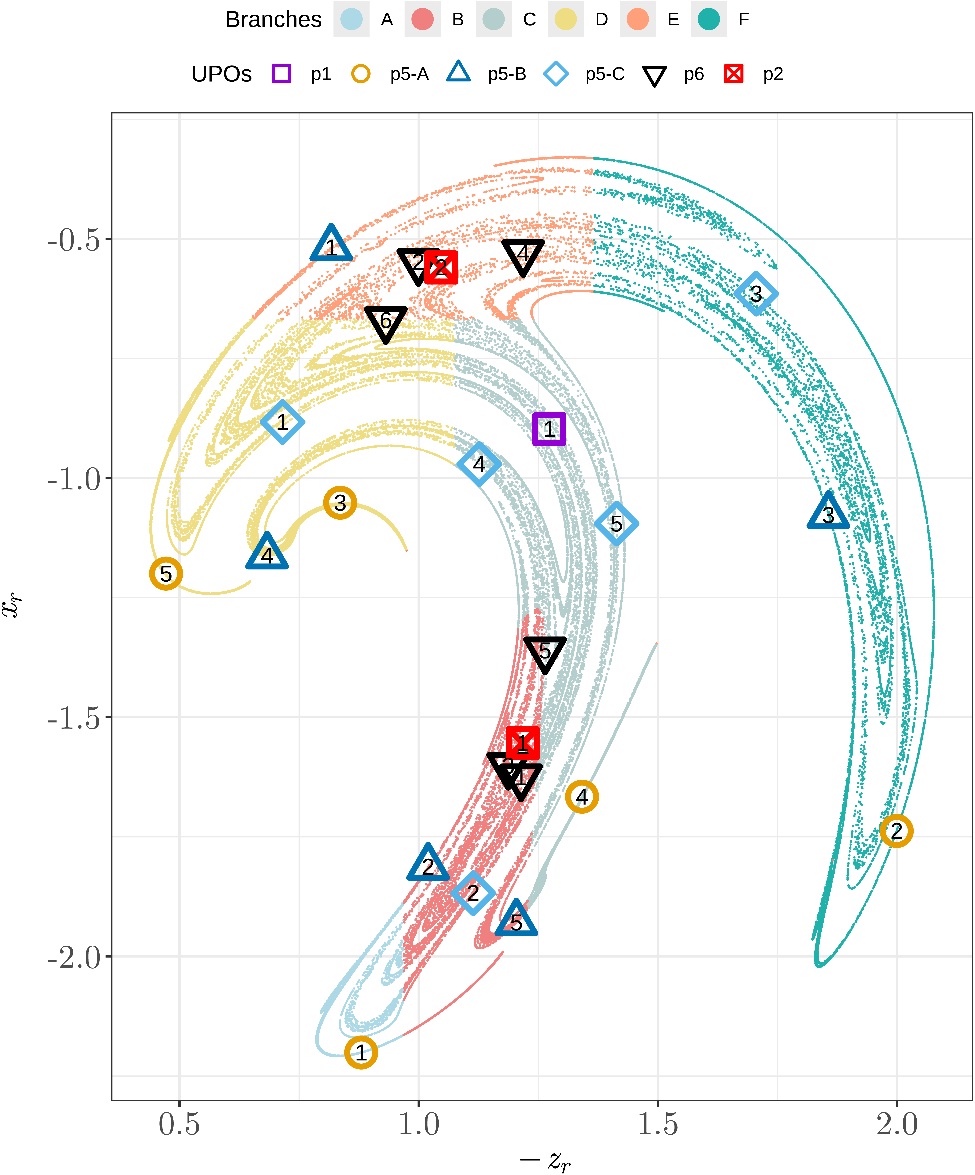}
    \caption{Poincaré section \eqref{eq:lorenz_02_section} with partition associated to the first return map build on $\gamma_n$ (Fig.~\ref{fig:appli_6branches}).}
    \label{fig:section_6branches}
\end{figure}

\begin{figure}
    \centering
    \includegraphics[width = .30\textwidth]{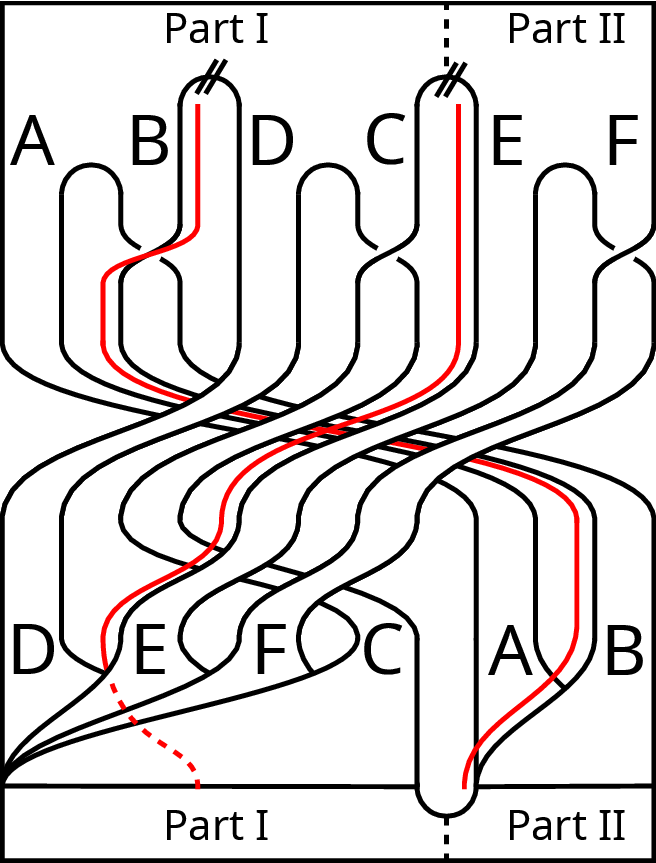}
    \caption{Reduced template of the Lorenz--84 attractor associated to the linking matrix \eqref{eq:linkingmatrix}.
    As a multidirectional stretching occurs, an additional partition is required to detail trajectories. The partition between part I and II is artificial because we build a Poincaré section with two main components ($\gamma_n$) \eqref{eq:gamma_n}. Because of the chosen splitting, the symbol of multidirectional stretching occurs twice to indicate that branches $D-C$ is the additional stretching direction. The red line corresponds to the p2 orbit located in the period--2 cavity encoded $BE$ (Tab.~\ref{tab:encoding}).}
    \label{fig:lorenz-84_template}
\end{figure}

Fig.~\ref{fig:section_6branches} details the partition of this first return map in six branches (Fig.~\ref{fig:appli_6branches}), which partition is also presented on the Poincaré section. 
Such a first return map with partition leads to a template with six branches where increasing branches have an even number of torsions and decreasing branches have an odd number of torsions. 
Using the previous plots (Figs.~\ref{fig:skeleton} and \ref{fig:template}) we propose a template (Fig.~\ref{fig:lorenz-84_template}) described by the linking matrix \eqref{eq:linkingmatrix}.
\begin{equation}
    \label{eq:linkingmatrix}
    \begin{array}{cc}
         & \begin{matrix}
        \;\;\;A & B & D & C & E & F 
         \end{matrix} \\
         \begin{matrix} A \\ B \\ D \\ C \\ E \\ F \end{matrix}
         & \left. \begin{bmatrix}
         \;0 & \;0  & 1  & 1  & 1 & 1 \\
         \;0 & 1 & 1 & 1 & 1 & 1 \\
         1 & 1 & \;0 & \;0 & \;0 & \;0 \\
         1 & 1 & \;0 & 1 & 1 & 1 \\
         1 & 1 & \;0 & 1 & \;0 & \;0 \\
         1 & 1 & \;0 & 1 & \;0 & 1 \\
         \end{bmatrix} \!\!\!\! \right|
    \end{array}
\end{equation}
This matrix details the torsions and the permutations between the six branches.
The branches of the templates are organized in a given order using a convention \cite{Tufillaro_1990}: the left to the right order corresponds to the bottom to top order. 
This convention is generally used for the insertion mechanism where the stretching and folding mechanism occurs.
Thus, we chose to also apply this convention for the three initial pairs of branches where $A$ and $B$ are on the left of $D$ and $C$ because they are below (Fig.~\ref{fig:section_6branches}); similarly, branches $D$ and $C$ are below branches $E$ and $F$.
Accordingly to this convention, from the relative organization of periodic points projected onto the $-z_r$-axis, this reordering of points induces negative crossings.

\begin{table}[htbp]
    \caption{Encoding of orbits according to partition of the first return map Fig.~\ref{fig:appli_6branches} where the number in column code corresponds to values in the symbols.}
    \centering
    \begin{tabular}{cccc}
        \hline \hline\\[-.3cm]
        Orbit & Symbol & Code & \quad Reduced code\quad\null \\
        \hline
        p1 & $\pun$ & $C_1$ & $C$ \\  
        p2 & $\pdeux$ & $B_1E_2$ & $BE$ \\  
        p5-A & $\pcinqa$ & $A_1F_2D_3C_4D_5$ & $AFD_3CD_5$ \\  
        p5-B & $\pcinqb$ & $E_1B_2F_3D_4B_5$ & $EB_2FDB_5$ \\  
        p5-C & $\pcinqc$ & $D_1B_2F_3C_4C_5$ & $DBFC_4C_5$ \\  
        p6 & $\psix$ & $B_1E_2B_3E_4C_5D_6$ & $B_1E_2B_3E_4CD$ \\  
        \hline
    \end{tabular}
    \label{tab:encoding}
\end{table}

Using this partition, we can encode periodic orbits with the symbols described in Table \ref{tab:encoding}.
To maintain links between the periodic orbits in the first return map (Fig.~\ref{fig:appli_6branches}) and the Poincaré section (Fig.~\ref{fig:section_6branches}), we conserve the position of the symbols as index if the symbol is duplicated (last column of Tab.~\ref{tab:encoding}).

\subsection{Theoretical linking numbers}

\begin{figure*}
    \centering
    \includegraphics[width = \textwidth]{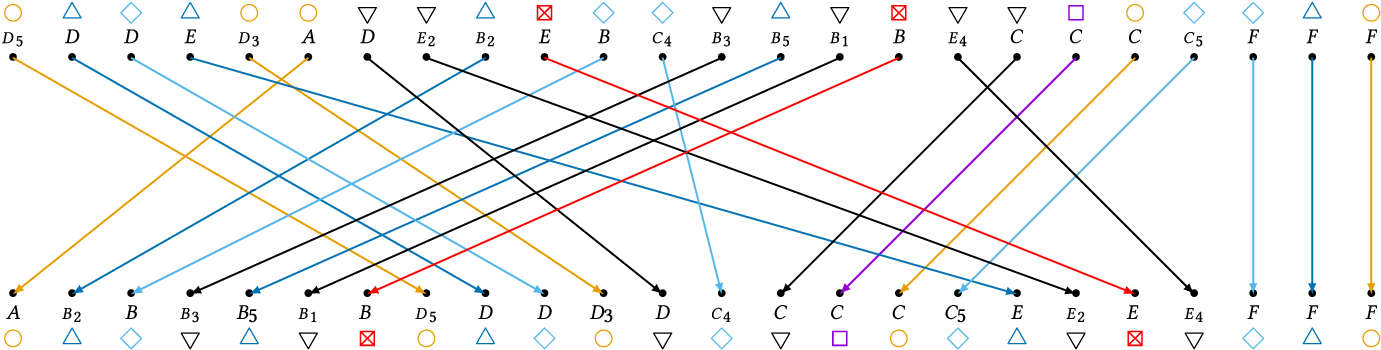}
    \caption{Negative crossings induced by the reordering of periodic points before template torsions and permutations (Fig.~\ref{fig:lorenz-84_template}).
    Upper order corresponds to the projection of the periodic points to $-z_r$-axis in order to preserve the orientation of the template from the inside to the outside (Fig.~\ref{fig:section_6branches}).
    Lower order corresponds to the order of this points according to the first-return map partition were branches of the templates are A-B-D-C-E-F (Fig.~\ref{fig:appli_6branches}).}
    \label{fig:minusz_axis}
\end{figure*}

To validate the reduced template shown in Figure \ref{fig:lorenz-84_template}, we have to prove that the theoretical linking numbers computed with the template correspond to the numerical ones Tab.~\ref{tab:lorenz84_02_lk}.
To consider negative crossings induced by the initial partition, the Figure ~\ref{fig:minusz_axis} details this reordering necessary to obtain a relative organization of periodic points for each branch of the template.
For instance, in Fig.~\ref{fig:section_6branches}, in the initial order using $-z_r$ values $E$ of p5-B orbits (upper triangle with 1) is on the left of $C$ of periodic one orbit. Considering the order with conventions, $C$ is on the left of $E$. As $D$ is higher (using $x_r$) the projection induces a negative crossing. 
Consequently, Fig.~\ref{fig:minusz_axis} details all the negative crossings induced by this partition of branches according to the convention named $N_\text{sep}$ for separation.

\begin{figure}[htb]
    \centering
    \begin{equation}
    \begin{array}{rl}
    lk(\pun, \psix) &= lk(C, B_1E_2B_3E_4CD) \\[1.1ex]
         & =\frac{1}{2} \left[ N_\text{sep} + 2M_{CB} + 2M_{CE} + M_{CC} + M_{CD} + N_{\text{ins}} \right] \\[1.1ex]
         & =\frac{1}{2} \left[ -2 + 2 + 2 + 1 + 0 + 3\right] = 3
    \end{array}
    \end{equation}
    \includegraphics[width = .35\textwidth]{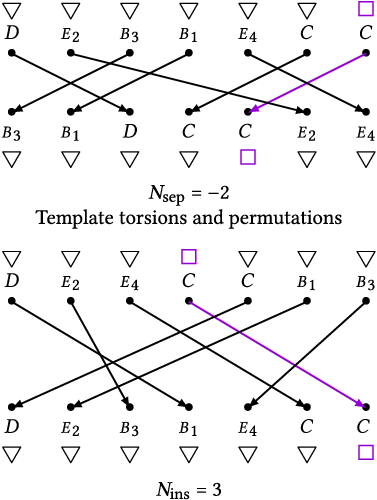}
    \caption{Theoretical linking number computation according to the phase space  
        $lk(\text{p1}, \text{p6}) = lk(\pun, \psix) = lk(C, B_1E_2B_3E_4CD) = 3$.
    }
    \label{fig:p1_p6}
\end{figure}

\begin{figure}[htb]
    \centering
    \begin{equation}
    \begin{array}{rl}
    lk(\pun, \psix) &= lk(C, B_1E_2B_3E_4CD) \\[1.1ex]
         & =\frac{1}{2} \left[2M_{CB} + 2M_{CE} + M_{CC} + M_{CD} + N_{\text{ins}} \right] \\[1.1ex]
         & =\frac{1}{2} \left[ 2 + 2 + 1 + 0 + 1\right] = 3
    \end{array}
    \end{equation}
    \includegraphics[width = .35\textwidth]{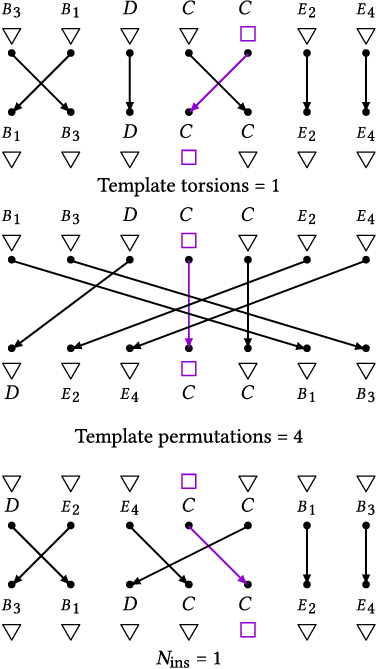}
    \caption{Theoretical linking number computation with the ordering of branches 
        $lk(\text{p1}, \text{p6}) = lk(\pun, \psix) = lk(C, B_1E_2B_3E_4CD) = 3$.
        For the other computation of this insertion number (Appendix \ref{app:calcul}), only the last graph is represented because torsions and permutations are directly computed using the linking matrix. For this computation torsions correspond to $M_{CC} = 1$ and permutations to $2M_{CB} + 2M_{CE} + M_{CD} = 4$.
    }
    \label{fig:p1_p6_alternative}
\end{figure}

To compute theoretical linking numbers happening in the phase space, this specific partition with two parts leads us to sum three categories of crossings. First, $N_\text{sep}$ crossings comes from the separation into six . 
Secondly, the crossings induced by the template according to the encoding. 
Lastly, $N_\text{ins}$ is the insertion number that contains positive crossings at the end of the template due to stretching and squeezing to the branch line.
The linking number is the half of the sum of all counted crossings \cite{Tufillaro_1990}.
The Figure \ref{fig:p1_p6} details the computation of the theoretical linking number between the period--1 and period--6 orbits.
The first crossing graph is a subgraph of Figure \ref{fig:minusz_axis} with the two orbits: $N_\text{sep} = -2$. 
The torsions and permutations are computed using the values in the linking matrix given by Eq. (\ref{eq:linkingmatrix}).
Finally, the positive crossings due to insertion are added after being calculated with the second graph (the transition from one symbol to the next one occurs at the end).
The initial separation of periodic points from the projection to the template (Fig.~\ref{fig:minusz_axis}) is required to explicit the relations of the UPOs in the phase space and in the template. 
However, this induces several negative crossings that are compensated by positive crossings during the squeezing mechanism at the end of the template (Fig.~\ref{fig:p1_p6}).
The template (Fig.~\ref{fig:lorenz-84_template}) is also valid by considering only the ordered projection of the periodic points (bottom order of Fig.~\ref{fig:minusz_axis}).
This is confirmed by the theoretical computation of the linking numbers with this second ordering of branches; the Figure \ref{fig:p1_p6_alternative} gives the details for period--1 and period--6 orbits.
In the classical framework (Fig.~\ref{fig:topological_characterization}), the encoding depends on parity and periodic points can be ordered using their encoding.
With the particularities of this attractor, we consider the order derived from the partition to confirm theoretical linking numbers.
In appendix \ref{app:calcul}, all numerical linking numbers of Table \ref{tab:lorenz84_02_lk} are confirmed by theoretical calculus using the two previously mentioned theoretical ways (phase space or theoretical) to compute them where $N_\text{sep} + N_\text{ins}$ in the computation using phase space corresponds to $N_\text{ins}$ in the theoretical framework.
The template (Fig~\ref{fig:lorenz-84_template} defined by Eq. (\ref{eq:linkingmatrix})) is thus validated to describe topological properties of the Lorenz-84 attractor $\mathcal{L}$.

This is a first template with approximations due to the thickness of the branches where the partition might be more accurate as well as the position of the periodic points for periodic orbits with a longer period.
Even if the reduced template is validated, some additional crossings occur in the squeezing mechanism of the theoretical computation of the linking numbers (Fig.~\ref{fig:p5a_p5b} branch $F$, Fig.~\ref{fig:p5a_p5c} branch $C$ and Fig.~\ref{fig:p5b_p5c} branch $F$).
These non-conventional crossings are due to the thickness of the reduced template approximate as for a dissipative system. They might also appear because of our ordering of points derived from the projection to $-z_{r}$ (Fig.~\ref{fig:minusz_axis}).
The reduced template (Fig.~\ref{fig:lorenz-84_template}) globally describes torsions and permutations of branches by underlying the multidirectional stretching mechanism of branches $A$ and $B$ over branches $D$ and $C$ : branches $B$ and $C$ have the same torsion number because they have the same behavior, but they end into two distinct parts of the attractor (part I and II).
The same mechanism occurs between the branches $D$ and $E$ because they come from two distinct parts, both ending into part I.
Branches $D$ and $C$ are located between the two holes corresponding to the (single) cavity, that is, where resides the period--2 orbit, they are thus associated to, respectively, the branches $E$ and $B$.
The first return map (Fig.~\ref{fig:lorenz84_02_section}b) could be split into four branches by grouping $B$ with $C$ and $D$ with $E$, this indicates that there are two simultaneous stretching and folding mechanisms. 
Further to some experiments on the Rössler system \cite{rosalie:tel-01143169, meneceur2025}, it appears that this mechanism is thus identifiable (Fig.~\ref{fig:lorenz84_02_section}b) in the first return maps even if the Poincaré section cannot be ideally positioned.

\section{Conclusion}
\label{sec:conclusion}

Using a colored tracer method, we first provide an overview of the structure organizing the flow of the Lorenz--84 attractor. This structure is built around a period--2 cavity that reveals a quite unusual case of toroidal chaos.  The analysis also reveals a multidirectional stretching rarely found in chaotic attractors. We conjecture that this mechanism is a necessary condition to develop a weakly dissipative chaos because it enables to generate and maintain the thickness of the flow. It was found possible to simplify its structure to a reduced skeleton, and then to a flattened skeleton --- that is, a two-dimensional branched manifold requiring a three dimensional space all along the flow --- involving seven branches revealing all the complexity of the attractor.

At the mean time, the more usual method developed for topological characterization was also applied: the periodic orbits were extracted, the linking numbers were computed and a template, that is, a two-dimensional branched manifold was generated from it. 
Here again, it was found possible to reduce the complexity of the template by considering the multidirectional stretching mechanism as the keystone of this chaotic system. 
It appears that, although no ideal Poincaré section were obtained to split the flow, the main structure of the chaotic mechanism, including the encoding of periodic orbits, enabled to detail the topological structure of the chaotic attractor.
The periodic-orbits were encoded with respect to the multidirectional stretching mechanism and a template was built starting from this specific mechanism then ending with a more classical stretching and squeezing mechanism. 
This template was validated by computing theoretically the linking numbers.

It was also tried to develop and validate a reduced template but the thickness of the attractor due to its weakly dissipative property may allow other templates.
In future works, the additional crossings obtained in insertion graphs could help new investigations to study the structure at the boundary of the inside cavity.

\section*{Conflict of Interest Statement}

The authors have no conflicts to disclose.

\section*{Data Availability Statement}

The data that support the findings of this study are available from the corresponding author upon reasonable request.

\printbibliography

\appendix

\section{Theoretical linking numbers}
\label{app:calcul}

The reduced template is validated because the theoretical linking computed hereinafter corresponds to the numerical linking numbers Tab~\ref{tab:lorenz84_02_lk}.
The following figures (Fig.~\ref{fig:p1_p5a} -- \ref{fig:p5b_p5c}) details the theoretical computation of the linking numbers according to the template defined by the linking matrix \eqref{eq:linkingmatrix}. 
\begin{itemize}
    \item $lk(\text{p1}, \text{p5-A}) = 2$, Fig.~\ref{fig:p1_p5a}
    \item $lk(\text{p1}, \text{p5-B}) = 2$, Fig.~\ref{fig:p1_p5b}
    \item $lk(\text{p1}, \text{p5-C}) = 2$, Fig.~\ref{fig:p1_p5c}
    \item $lk(\text{p1}, \text{p6}) = 3$, Fig.~\ref{fig:p1_p6} and Fig.~\ref{fig:p1_p6_alternative}
    \item $lk(\text{p1}, \text{p2}) = 1$, Fig.~\ref{fig:p1_p2}
    \item $lk(\text{p5-A}, \text{p5-B}) = 10$, Fig.~\ref{fig:p5a_p5b}
    \item $lk(\text{p5-A}, \text{p5-C}) = 10$, Fig.~\ref{fig:p5a_p5c}
    \item $lk(\text{p5-B}, \text{p5-C}) = 10$, Fig.~\ref{fig:p5b_p5c}
\end{itemize}

\begin{figure*}[b]
    \centering
    \begin{tabular}{c|c}
    $
    \begin{aligned}
        lk(\pun, \pcinqa)&= lk(C, AFD_3CD_5)  \\[1.1ex]
         & =\frac{1}{2} \left[ N_\text{sep} + M_{CA} + M_{CF} + 2M_{CD} + M_{CC} + N_{\text{ins}} \right] \\[1.1ex]
         & =\frac{1}{2} \left[ 0 + 1 + 1 + 0 + 1 + 1\right] = 2
    \end{aligned}
    $
    & 
    $
    \begin{aligned}
        lk(\pun, \pcinqa)&= lk(C, AFD_3CD_5)  \\
         & =\frac{1}{2} \left[ M_{CA} + M_{CF} + 2M_{CD} + M_{CC} + N_{\text{ins}} \right] \\[1.1ex]
         & =\frac{1}{2} \left[ 1 + 1 + 0 + 1 + 1\right] = 2
    \end{aligned}
    $
    \\
    \includegraphics[width = .3\textwidth]{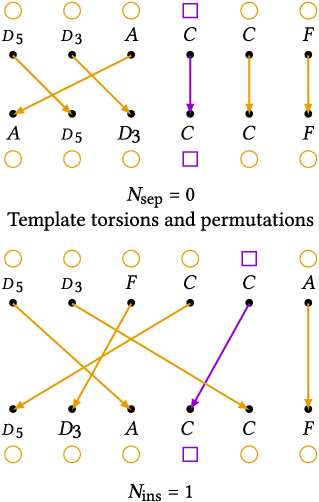}
    & 
    \includegraphics[width = .3\textwidth]{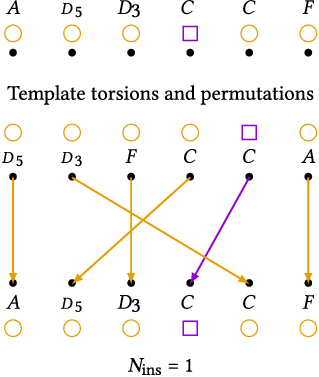}
    \end{tabular}
    \caption{Theoretical linking number computation  
        $lk(\text{p1}, \text{p5-A}) = lk(\pun, \pcinqa) = lk(C, AFD_3CD_5) = 2$.
    }
    \label{fig:p1_p5a}
\end{figure*}

\begin{figure*}[b]
    \centering
    \begin{tabular}{c|c}
    $\begin{array}{rl}
        lk(\pun, \pcinqb)&= lk(C, EB_2FDB_5)  \\[1.1ex]
         & =\frac{1}{2} \left[ N_\text{sep} + M_{CE} + 2M_{CB} + M_{CF} + M_{CD} + N_{\text{ins}} \right] \\[1.1ex]
         & =\frac{1}{2} \left[ -1 + 1 + 2 + 1 + 0 + 1\right] = 2
    \end{array}$ &
    $\begin{array}{rl}
        lk(\pun, \pcinqb)&= lk(C, EB_2FDB_5)  \\[1.1ex]
         & =\frac{1}{2} \left[ M_{CE} + 2M_{CB} + M_{CF} + M_{CD} + N_{\text{ins}} \right] \\[1.1ex]
         & =\frac{1}{2} \left[ 1 + 2 + 1 + 0 + 0\right] = 2
    \end{array}$ \\
    \includegraphics[width = .3\textwidth]{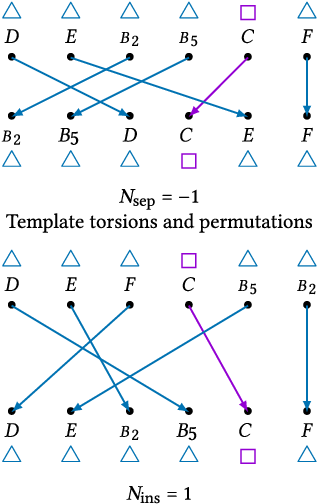}
    &
    \includegraphics[width = .3\textwidth]{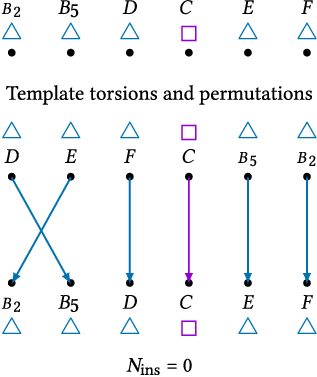}
    \end{tabular}
    \caption{Theoretical linking number computation  
        $lk(\text{p1}, \text{p5-B}) = lk(\pun, \pcinqb) = lk(C, EB_2FDB_5) = 2$.
    }
    \label{fig:p1_p5b}
\end{figure*}

\begin{figure*}
    \centering
    \begin{tabular}{c|c}
    $\begin{array}{rl}
        lk(\pun, \pcinqc)&= lk(C, DBFC_4C_5)  \\[1.1ex]
         & =\frac{1}{2} \left[N_\text{sep} + M_{CD} + M_{CB} + M_{CF} + 2M_{CC} + N_{\text{ins}} \right] \\[1.1ex]
         & =\frac{1}{2} \left[ 0 + 0 + 1 + 1 + 2 + 0 \right] = 2
    \end{array}$ & 
    $\begin{array}{rl}
        lk(\pun, \pcinqc)&= lk(C, DBFC_4C_5)  \\[1.1ex]
         & =\frac{1}{2} \left[ M_{CD} + M_{CB} + M_{CF} + 2M_{CC} + N_{\text{ins}} \right] \\[1.1ex]
         & =\frac{1}{2} \left[ 0 + 1 + 1 + 2 + 0 \right] = 2
    \end{array}$ \\
    \includegraphics[width = .3\textwidth]{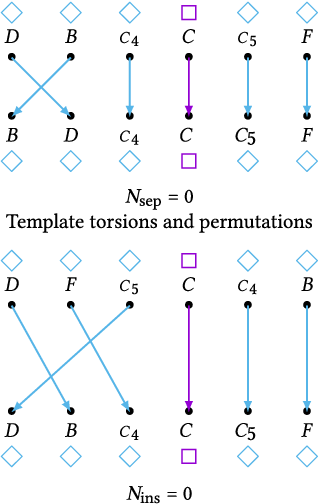} & 
    \includegraphics[width = .3\textwidth]{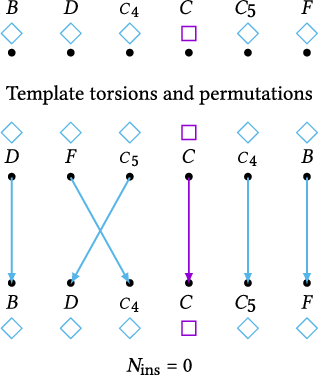}
    \end{tabular}
    \caption{Theoretical linking number computation  
        $lk(\text{p1}, \text{p5-C}) =  lk(\pun, \pcinqc) = lk(C, DBFC_4C_5) = 2$.
    }
    \label{fig:p1_p5c}
\end{figure*}

\begin{figure*}
    \centering
    \begin{tabular}{c|c}
    $\begin{array}{rl}
        lk(\pun, \pdeux)&= lk(C, EB)  \\[1.1ex]
         & =\frac{1}{2} \left[ N_\text{sep} + M_{CE} + M_{CB} + N_{\text{ins}} \right] \\[1.1ex]
         & =\frac{1}{2} \left[ -1 + 1 + 1 + 1 \right] = 1
    \end{array}$ &
    $\begin{array}{rl}
        lk(\pun, \pdeux)&= lk(C, EB)  \\[1.1ex]
         & =\frac{1}{2} \left[ M_{CE} + M_{CB} + N_{\text{ins}} \right] \\[1.1ex]
         & =\frac{1}{2} \left[ 1 + 1 + 0 \right] = 1
    \end{array}$ \\
    \includegraphics[width = .3\textwidth]{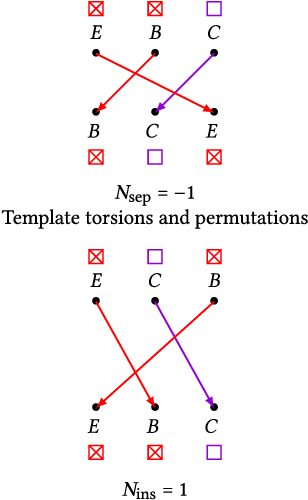} & 
    \includegraphics[width = .3\textwidth]{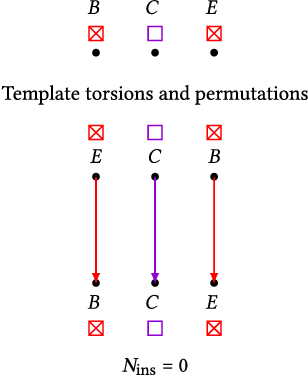}
    \end{tabular}
    \caption{Theoretical linking number computation  
        $lk(\text{p1}, \text{p2}) =  lk(\pun, \pdeux) = lk(C, EB) = 1 $.
    }
    \label{fig:p1_p2}
\end{figure*}

\begin{figure*}
    \centering
    \begin{tabular}{c|c}
    $\begin{array}{rl}
     lk(\pcinqa, \pcinqb) &= lk(AFD_3CD_5, EB_2FDB_5) \\
         & =\frac{1}{2} \left[ N_\text{sep} + N_{\text{ins}} + \right. \\
         & \qquad M_{AE} + 2M_{AB} + M_{AF} + M_{AD} + \\
         & \qquad M_{FE} + 2M_{FB} + M_{FF} + M_{FD} + \\
         & \qquad 2 \left( M_{DE} + 2M_{DB} + M_{DF} + M_{DD} \right) + \\
         & \left. \qquad M_{CE} + 2M_{CB} + M_{CF} + M_{CD} \right] \\
         & =\frac{1}{2} \left[ -8 + 14 + 14 \right] = 10
    \end{array}$ & 
    $\begin{array}{rl}
     lk(\pcinqa, \pcinqb) &= lk(AFD_3CD_5, EB_2FDB_5) \\
         & =\frac{1}{2} \left[ N_{\text{ins}} + \right. \\
         & \qquad M_{AE} + 2M_{AB} + M_{AF} + M_{AD} + \\
         & \qquad M_{FE} + 2M_{FB} + M_{FF} + M_{FD} + \\
         & \qquad 2 \left( M_{DE} + 2M_{DB} + M_{DF} + M_{DD} \right) + \\
         & \left. \qquad M_{CE} + 2M_{CB} + M_{CF} + M_{CD} \right] \\
         & =\frac{1}{2} \left[ 14 + 6 \right] = 10
    \end{array}$
    \\
    \includegraphics[width = .4\textwidth]{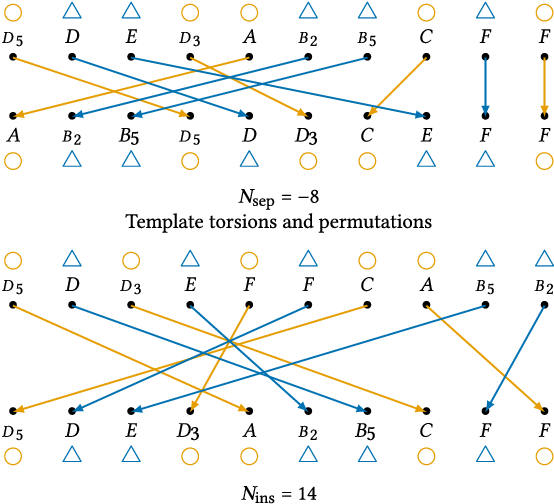} & 
    \includegraphics[width = .4\textwidth]{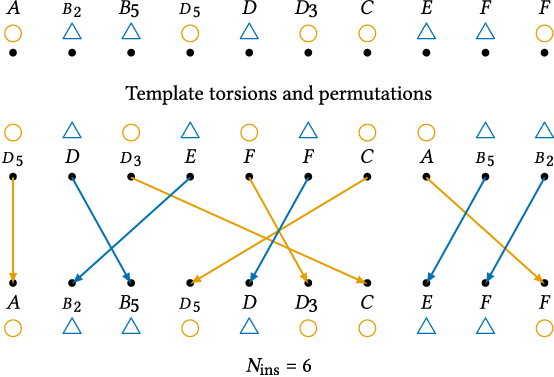}
    \end{tabular}
    \caption{Theoretical linking number computation  
        $lk(\text{p5-A}, \text{p5-B}) =  lk(\pcinqa, \pcinqb) = lk(AFD_3CD_5, EB_2FDB_5) = 10 $.
    }
    \label{fig:p5a_p5b}
\end{figure*}

\begin{figure*}
    \centering
    \begin{tabular}{c|c}
    $\begin{array}{rl}
     lk(\pcinqa, \pcinqc) &= lk(AFD_3CD_5, DBFC_4C_5) \\
         & =\frac{1}{2} \left[ N_\text{sep} + N_{\text{ins}} + \right. \\
         & \qquad M_{AE} + 2M_{AB} + M_{AF} + M_{AD} + \\
         & \qquad M_{FE} + 2M_{FB} + M_{FF} + M_{FD} + \\
         & \qquad 2 \left( M_{DE} + 2M_{DB} + M_{DF} + M_{DD} \right) + \\
         & \left. \qquad M_{CE} + 2M_{CB} + M_{CF} + M_{CD} \right] \\
         & =\frac{1}{2} \left[ -3 + 14 + 9 \right] = 10
    \end{array}$ & 
    $\begin{array}{rl}
     lk(\pcinqa, \pcinqc) &= lk(AFD_3CD_5, DBFC_4C_5) \\
         & =\frac{1}{2} \left[ N_{\text{ins}} + \right. \\
         & \qquad M_{AD} + M_{AB} + M_{AF} + 2M_{AC} + \\
         & \qquad M_{FD} + M_{FB} + M_{FF} + 2M_{FC} + \\
         & \qquad 2 \left( M_{DD} + M_{DB} + M_{DF} + 2M_{DC} \right) + \\
         & \left. \qquad M_{CD} + M_{CB} + M_{CF} + 2M_{CC} \right] \\
         & =\frac{1}{2} \left[ 14 + 6 \right] = 10
    \end{array}$
    \\
    \includegraphics[width = .4\textwidth]{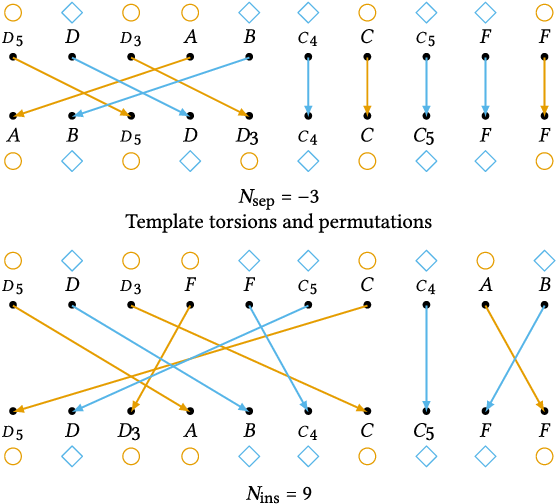} & 
    \includegraphics[width = .4\textwidth]{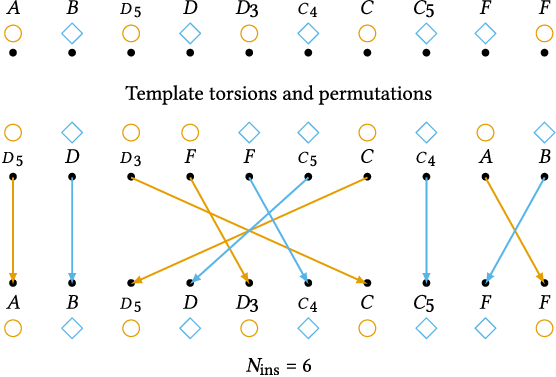}
    \end{tabular}
    \caption{Theoretical linking number computation  
        $lk(\text{p5-A}, \text{p5-C}) =  lk(\pcinqa, \pcinqc) = lk(AFD_3CD_5, DBFC_4C_5) = 10 $.
    }
    \label{fig:p5a_p5c}
\end{figure*}

\begin{figure*}
    \centering
    \begin{tabular}{c|c}
    $\begin{array}{rl}
     lk(\pcinqa, \pcinqc) &= lk(EB_2FDB_5, DBFC_4C_5) \\
         & =\frac{1}{2} \left[ N_\text{sep} + N_{\text{ins}} + \right. \\
         & \qquad M_{DE} + 2M_{DB} + M_{DF} + M_{DD} + \\
         & \qquad M_{BE} + 2M_{BB} + M_{BF} + M_{BD} + \\
         & \qquad  M_{FE} + 2M_{FB} + M_{FF} + M_{FD}  + \\
         & \left. \qquad 2 \left( M_{CE} + 2M_{CB} + M_{CF} + M_{CD}\right) \right] \\
         & =\frac{1}{2} \left[ -7 + 18 + 9 \right] = 10
    \end{array}$ & 
    $\begin{array}{rl}
     lk(\pcinqa, \pcinqc) &= lk(EB_2FDB_5, DBFC_4C_5) \\
         & =\frac{1}{2} \left[ N_{\text{ins}} + \right. \\
         & \qquad M_{DE} + 2M_{DB} + M_{DF} + M_{DD} + \\
         & \qquad M_{BE} + 2M_{BB} + M_{BF} + M_{BD} + \\
         & \qquad  M_{FE} + 2M_{FB} + M_{FF} + M_{FD}  + \\
         & \left. \qquad 2 \left( M_{CE} + 2M_{CB} + M_{CF} + M_{CD}\right) \right] \\
         & =\frac{1}{2} \left[ 18 + 2 \right] = 10
    \end{array}$
    \\
    \includegraphics[width = .4\textwidth]{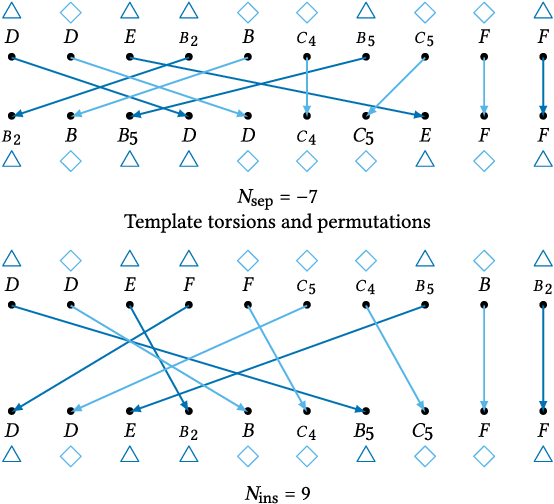} & 
    \includegraphics[width = .4\textwidth]{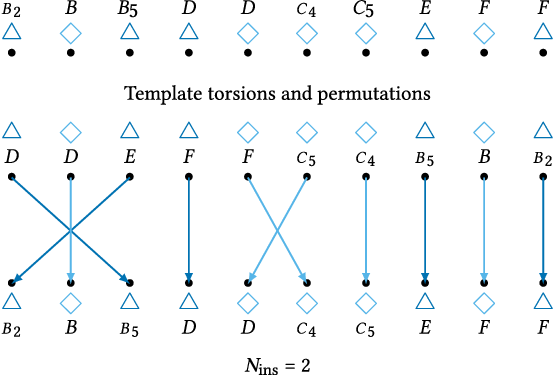}
    \end{tabular}
    \caption{Theoretical linking number computation  
        $lk(\text{p5-B}, \text{p5-C}) =  lk(\pcinqb, \pcinqc) = lk(EB_2FDB_5, DBFC_4C_5) = 10 $.
    }
    \label{fig:p5b_p5c}
\end{figure*}

Similarly, this methodology is apply to compute the other theoretical linking numbers (Tab.~\ref{tab:nins_nsep} contains values of the separation and insertion graphs).

$\begin{aligned}
     lk(\text{p5-A}, \text{p2}) &= lk(AFD_3CD_5, BE) \\
         & =\frac{1}{2} \left[ N_{\text{ins}} + \right. \\
         & \qquad M_{AE} + M_{AB}  + \\
         & \qquad M_{FE} + M_{FB}  + \\
         & \qquad 2 \left( M_{DE} + M_{DB}  \right) + \\
         & \left. \qquad M_{CE} + M_{CB}  \right] \\
         & =\frac{1}{2} \left[ 2 + 6 \right] = 4
\end{aligned}$

$\begin{aligned}
     lk(\text{p5-B}, \text{p2}) &= lk(EB_2FDB_5, BE) \\
         & =\frac{1}{2} \left[ N_{\text{ins}} + \right. \\
         & \qquad M_{EE} + M_{EB}  + \\
         & \qquad 2 \left( M_{BE} + M_{BB}  \right) + \\
         & \qquad M_{FE} + M_{FB}  + \\
         & \left. \qquad M_{DE} + M_{DB}  \right] \\
         & =\frac{1}{2} \left[ 1 + 7 \right] = 4
\end{aligned}$

$\begin{aligned}
     lk(\text{p5-C}, \text{p2}) &= lk(DBFC_4C_5, BE) \\
         & =\frac{1}{2} \left[ N_{\text{ins}} + \right. \\
         & \qquad M_{DE} + M_{EB}  + \\
         & \qquad M_{BE} + M_{BB}  + \\
         & \qquad M_{FE} + M_{FB}  + \\
         & \left. \qquad 2 \left( M_{CE} + M_{CB}  \right) \right] \\
         & =\frac{1}{2} \left[ 0 + 8 \right] = 4
    \end{aligned}$
    
    $\begin{aligned}
     lk(\text{p6}, \text{p2}) &= lk(B_1E_2B_3E_4CD, BE) \\
         & =\frac{1}{2} \left[ N_{\text{ins}} + \right. \\
         & \qquad 2 (M_{BB} + M_{BE})  + \\
         & \qquad 2 (M_{EB} + M_{EE})  + \\
         & \qquad    M_{CB} + M_{CE}  + \\
         & \left. \qquad M_{DB} + M_{DE} \right] \\
         & =\frac{1}{2} \left[ 1 + 9 \right] = 5
    \end{aligned}$

    $\begin{aligned}
     lk(\text{p5-A}, \text{p6}) &= lk(AFD_3CD_5, B_1E_2B_3E_4CD) \\
         & =\frac{1}{2} \left[ N_{\text{ins}} + \right. \\
         & \qquad 2 M_{AB} + 2M_{AE} + M_{AC} + M_{AD} + \\
         & \qquad 2 M_{FB} + 2M_{FE} + M_{FC} + M_{FD} + \\
         & \qquad 2 \left( 2M_{DB} + 2M_{DE} + M_{DC} + M_{DD}\right) + \\
         & \left. \qquad 2 M_{CB} + 2M_{CE} + M_{CC} + M_{CD} \right] \\
         & =\frac{1}{2} \left[ 8 + 16 \right] = 12
    \end{aligned}$
    
        $\begin{aligned}
     lk(\text{p5-B}, \text{p6}) &= lk(EB_2FDB_5, B_1E_2B_3E_4CD) \\
         & =\frac{1}{2} \left[ N_{\text{ins}} + \right. \\
         & \qquad 2 M_{EB} + 2M_{EE} + M_{EC} + M_{ED} + \\
         & \qquad 2 \left( 2M_{BB} + 2M_{BE} + M_{BC} + M_{BD}\right) + \\
         & \qquad 2 M_{FB} + 2M_{FE} + M_{FC} + M_{FD} + \\
         & \left. \qquad 2 M_{DB} + 2M_{DE} + M_{DC} + M_{DD} \right] \\
         & =\frac{1}{2} \left[ 4 + 20 \right] = 12
    \end{aligned}$
    
        $\begin{aligned}
     lk(\text{p5-C}, \text{p6}) &= lk(DBFC_4C_5, B_1E_2B_3E_4CD) \\
         & =\frac{1}{2} \left[ N_{\text{ins}} + \right. \\
         & \qquad 2 M_{DB} + 2M_{DE} + M_{DC} + M_{DD} + \\
         & \qquad 2 M_{BB} + 2M_{BE} + M_{BC} + M_{BD} + \\
         & \qquad 2 M_{FB} + 2M_{FE} + M_{FC} + M_{FD} + \\
         & \left. \qquad 2 \left( 2 M_{CB} + 2M_{CE} + M_{CC} + M_{CD} \right) \right] \\
         & =\frac{1}{2} \left[ 3 + 21 \right] = 12
    \end{aligned}$
    
\begin{table}[htbp]
    \centering
    \caption{Values of the separation and insertions graphs used to compute theoretical linking numbers.}
    \label{tab:nins_nsep}
    \begin{tabular}{cc|cc|c}
        & & Phase & space & Theoretical \\
        Orbit 1 & Orbit 2 & $N_\text{sep}$ & $N_\text{ins}$  & $N_\text{ins}$ \\ \hline
        p5-A & p2 & -3 & 5 & 2    \\
        p5-B & p2 & -3 & 4 & 1    \\
        p5-C & p2 & -5 & 5 & 0    \\
        p6 & p2 & -5 & 6 & 1    \\
        p5-A & p6 & -6 & 14 & 8    \\
        p5-B & p6 & -10 & 14 & 4    \\
        p5-C & p6 & -9 & 12 & 3    \\
    \end{tabular}
\end{table}

\end{document}